\documentclass[unsortedaddress,eqsecnum,aps]{revtex4}  
\usepackage[dvips]{graphicx}
\usepackage{longtable}
\usepackage{bm}
\usepackage{color}
\usepackage{amsmath}
\usepackage{amsfonts}
\usepackage{relsize}
\usepackage{booktabs}
\usepackage{amssymb}
\newcommand{\singlespace}{\tiny\renewcommand{\baselinestretch}{1.0}\normalsize}

\begin{document}


\title{Wavelet-Based Linear-Response Time-Dependent Density-Functional Theory}

\author{Bhaarathi Natarajan$^{a,b}$}
\email[]{Bhaarathi.Natarajan@UJF-Grenoble.FR}
\author{Luigi Genovese$^{b}$}
\email[]{Luigi.Genovese@ESRF.FR}
\author{Mark E.\ Casida$^a$}
\email[]{Mark.Casida@UJF-Grenoble.FR}
\author{Thierry Deutsch$^{b}$}
\email[]{Thierry.Deutsch@CEA.FR}
\author{Olga N.\ Burchak$^{a}$}
\email[]{olga_burchak@yahoo.fr}
\author{Christian Philouze$^{a}$}
\email[]{Christian.Philouze@ujf-grenoble.fr}
\author{Maxim Y.\ Balakirev$^{c}$}
\email[]{maxim.balakirev@cea.fr}
\affiliation{
        $^a$ Laboratoire de Chimie Th\'eorique,
        D\'epartement de Chimie Mol\'ecularie (DCM, UMR CNRS/UJF 5250),
        Institut de Chimie Mol\'eculaire de Grenoble (ICMG, FR2607),
        Universit\'e Joseph Fourier (Grenoble I),
        301 rue de la Chimie, BP 53,
        F-38041 Grenoble Cedex 9, France\\
        $^b$ UMR-E CEA/UJF-Grenoble 1, 
             INAC, Grenoble, F-38054, France\\
        $^c$ iRTSV/Biopuces, 17 rue des Martyrs, 
        38 054 Grenoble Cedex 9, France}


\begin{abstract}

Linear-response time-dependent (TD) density-functional theory (DFT) has been implemented in
the pseudopotential wavelet-based electronic structure program {\sc BigDFT} and results are compared against
those obtained with the all-electron Gaussian-type orbital program {\sc deMon2k} for the calculation of
electronic absorption spectra of N$_2$ using the TD local density approximation (LDA).
The two programs give comparable excitation energies and absorption spectra once suitably extensive basis sets are used.  
Convergence of LDA density orbitals and orbital energies to the basis-set limit is significantly
faster for {\sc BigDFT} than for {\sc deMon2k}.  However the number of virtual orbitals used in
TD-DFT calculations is a parameter in {\sc BigDFT}, while all virtual orbitals are included in 
TD-DFT calculations in {\sc deMon2k}.  As a reality check, we report the x-ray crystal 
structure and the measured and calculated absorption spectrum (excitation energies and oscillator
strengths) of the small organic molecule $N$-cyclohexyl-2-(4-methoxyphenyl)imidazo[1,2-$a$]pyridin-3-amine. 

\end{abstract}

\maketitle


\section{Introduction}
\label{sec:intro}

The last century witnessed the birth, growth, and increasing awareness of the importance of
quantum mechanics for describing the behavior of electrons in atoms, molecules, and solids.  One key
to unlocking the door to widespread applications was the advent of scientific computing in the
1960s.  New computational methods had to be developed --- and continue to be developed --- to make
use of the increasing computational power.  Wavelet-based
methods are a comparative newcomer to the world of electronic structure algorithms (Refs.~\cite{A99,G98}
provide useful, if dated, reviews) but offer
increasing accuracy for grid-based density-functional theory (DFT) calculations of molecular
properties.  Although DFT is a theory of the ground stationary state, excited states may be
treated by the complementary time-dependent (TD) theory.  This paper reports the first implementation 
of TD-DFT in a wavelet-based code --- namely our implementation
of linear response TD-DFT in the {\sc BigDFT} \cite{bigdft} program. 
TD-DFT results obtained with our implementation in {\sc BigDFT} and with the Gaussian-type orbital (GTO)
based program {\sc deMon2k} \cite{deMon2k} are compared and {\sc BigDFT} is used to calculate and analyse
the absorption spectrum of a recently-synthesized small organic molecule \cite{BMO11}.

Modern computing requires discretization and this has been historically treated in DFT
and non-DFT {\em ab initio} \cite{abinitio} electronic structure calculations by basis set expansions.  
Because the basis set is necessarily finite, electronic structure algorithms for solids and 
molecules have developed as refinements on physically-sensible zero-order systems, based 
upon the reasonable idea that fewer basis functions should be needed when the underlying physics is 
already approximately correct.  Thus computations on solids used plane-wave
basis sets partly because of analogies with the free-electron model of metals and partly
for other reasons (ease of integral evaluation and Bloch's theorem taking explicit account of
crystal symmetry.)  Similarly computations
on molecules are typically based upon the linear combination of atomic orbital (LCAO) approximation
to molecular orbitals, with split-valence, polarization, and diffuse functions added as needed to
go beyond this crude first approximation.  A key feature of finite-basis-set calculations has been 
the use of the variational principle in order to guarantee smooth convergence from above to the true 
ground-state energy.

Meanwhile a different set of methods developed within the larger context of engineering applications, 
where model-independent methods are important for treating a variety of complex systems.
These methods involve spatial discretization over a grid of points and include finite-difference methods, 
finite-element methods, and (more recently) wavelet methods.  Use of these 
methods in solid-state electronic structure calculations is particularly natural in areas such as materials 
science where physics meets engineering \cite{CKSL08}, in real-time applications \cite{CMA+06}
and also in chemical engineering \cite{MJ94}.

Although one of the first applications of computers to scientific problems was to solve the Schr\"odinger
equation by direct numerical integration using the Cooley-Cashion-Zare approach \cite{C61,C63,Z64} 
and this method soon became the preferred way to find atomic orbitals \cite{F72,M73,F73}, quantum chemists 
have been slow to accept grid-based methods.  There have 
been at least three reasons for this.  The first is the principle that any two quantum chemistry programs 
should give the same answer to machine precision when the same calculation is performed.  Historically
this principle has been essential for debugging and assuring the consistency of the multiple sophisticated 
GTO-based programs common in quantum 
chemistry.  The second reason is the fear that numerical noise would undermine the variational principle
and make chemical accuracy (1 kcal/mol) impossible to achieve for realistic chemical problems.  The
first principle has gradually been abandoned with the wide-spread adaption in quantum chemistry of DFT 
and its associated grid-based algorithms.  The second reason is simply an obsolete fear as grid-based
methods are now more precise and free of numerical error than ever before.
A third reason for the slow acceptance of grid-based methods in quantum chemistry is the
conviction that the continuum-like unoccupied orbitals typically produced by grid-based methods
are not the most efficient orbital-basis set to use when large configuration interaction (CI) or 
many-body perturbation theory (MBPT) type expansions are needed for the accurate description of 
electron correlation.   Rather it is more useful to replace continuum-like unoccupied orbitals
with natural orbitals or with other orbitals localized in the same region of space as the
occupied orbitals since this is where electron correlation occurs.  We discuss this problem
further in Sec.~\ref{sec:validation}.

Wavelet-based codes represent some of the latest evolutions in grid-based methods for the problem
of electronic structure calculations.  Daubechies classic 1992 book \cite{D92} helped to establish 
wavelet theory as the powerful, flexible, and still-growing toolbox we know today.
Wavelet theory has deep roots in applied mathematics and computational theory, and has benefited 
from the various points of view and expertises of researchers in markedly different but complementary 
disciplines. (Ingrid Daubechies describes this aspect of wavelet theory rather well in a 1996 viewpoint 
article \cite{D96}.)  Wavelet theory somewhat resembles the older Fourier theory in its use of continuous 
and discrete transforms and underlying grids.  However wavelet theory is designed to embody the powerful idea of
multiresolution analysis (MRA) that coarse features are large-scale objects while fine-scale 
features tend to be more localized.  Computationally, a low resolution description is afforded
by a set of so-called ``scaling functions'' placed at nodes of a coarse spatial grid.  Note that
scaling functions are typically only placed where they are needed on this grid.  Wavelets or 
``detail functions'' are then added adaptively on a finer grid in regions of space where more resolution is needed.
This provides yet another attractive feature for grid-based calculations, namely  
flexible boundary conditions.
A subtler, and also highly desirable feature, is that the coefficients of the scaling and wavelet
expansions are of comparable size and so properly reflect an even balance between different
length scales.  The reader interested in more information about wavelets is referred to the
classic book of Daubechies \cite{D92} and for applications in theoretical physics
is referred to Refs.~\cite{G98,GM10}.

The first applications of wavelet theory to solving the Schr\"odinger equation 
began in the mid-1990s \cite{A99}. (Interest was also expressed by quantum chemists at around the same 
time~\cite{FD93,C96b}.)  At least two important wavelet-based codes have been developed for
solving the Kohn-Sham equation of the traditional Hohenberg-Kohn-Sham ground-state DFT \cite{HK64,KS65}.
These are {\sc Madness} \cite{HFYB03} and {\sc BigDFT} \cite{bigdft}.  This paper concerns {\sc BigDFT}.

Since Hohenberg-Kohn-Sham DFT is a ground-state theory, a different theory is needed to treat electronic excited states and,
in particular, to calculate absorption spectra.  The TD-DFT formalism \cite{RG84} complements that
of DFT by laying the ground work for calculating the time-dependent response of the charge density to
an applied perturbation.  Excitation spectra may then be calculated through linear response theory 
using, for example, the equations developed by one of us \cite{C95}.  Implementing these equations
in a wavelet code is not entirely straightforward since integral evalulation is performed differently
than in traditional TD-DFT codes.  Thus a key point in the present paper is how we handle integral evaluation
in our implementation of TD-DFT in {\sc BigDFT}. 

\begin{figure}
\includegraphics[angle=0,width=0.3\textwidth]{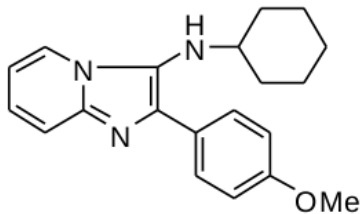}
\caption{$N$-cyclohexyl-2-(4-methoxyphenyl)imidazo[1,2-$a$]pyridin-3-amine (Flugi {\bf 6}). \label{fig:Olga_Flugi2} }
\end{figure}

Our implementation is first validated by comparison against TD-DFT calculations with the GTO-based
program {\sc deMon2k} for the historically-important test case of N$_2$ \cite{JCS96,CJC+98}.  This allows
us to see and discuss some of the pros and cons of wavelets versus GTOs.  We then go on to apply the
method to a real-world application, namely the calculation of the absorption spectrum of a molecule
of potential interest as a fluorescent probe in biological applications.  This molecule,
$N$-cyclohexyl-2-(4-methoxyphenyl)imidazo[1,2-$a$]pyridin-3-amine (Fig.~\ref{fig:Olga_Flugi2}), will simply
be referred to as Flugi {\bf 6} \cite{BMO11}. Since this molecule has not been thoroughly characterized 
before we have included an experimental section describing our determination of its x-ray crystal structure.

This paper is organized as follows: The basic equations of TD-DFT are reviewed in the next section.
In Sec.~\ref{sec:BigDFT}, we briefly review the idea of wavelets, explain how we handle the integral
evaluation in our implementation of TD-DFT, and Section~\ref{sec:validation} discuss the pros and cons of the wavelet implementation
in the context of N$_2$ calculations with {\sc BigDFT} and the GTO-based code {\sc deMon2k}.  
Section~\ref{sec:flugi6} reports the x-ray crystal structure, and experimental and calculated
UV/visible absorption spectrum of Flugi {\bf 6}.  An assignment is also made of the peaks appearing
in the spectrum.  The final section contains our concluding discussion.
 
\section{Time-Dependent Density-Functional Theory}
\label{sec:TDDFT}

As evidenced by the numerous review articles 
\cite{GK90,GUG94,C95,GDP96,C96,BG98,L01,VBB+01,C02,D03,MG03,MG04,BWG05,DH05,CMA+06,C08,C09,CH12}
and books \cite{MS90,MUN+06,GMNR11} written on the subject, TD-DFT is now such a well-established
formalism that little additional review seems necessary.  Nevertheless the purpose of this
section is to recall a few key equations in order to keep the present paper reasonably self-contained
and to introduce notation.  

Time-dependent DFT builds upon the Kohn-Sham formulation of ground-state DFT \cite{KS65}.
In its modern formulation, Kohn-Sham DFT is spin-DFT so that the total energy involves
an exchange-correlation (xc) energy which depends upon the density $\rho_\alpha$ of spin-$\alpha$
electrons and the density $\rho_{\ensuremath{\mathsmaller\beta}}$ of spin-$\beta$ electrons.  These are obtained as
the sum of the densities of the occupied orbitals of each spin,
\begin{equation}
  \rho_\sigma({\bf r}) = e \sum_p n_{p\sigma} |\psi_{p\sigma}({\bf r}) |^2 \, ,
  \label{eq:tddft.1}
\end{equation}
where $n_{p\sigma}$ is the occupation number of orbital $\psi_{p\sigma}$ and,
the total charge density is given as the sum of the charge densities of two spin components.
The Kohn-Sham orbitals are obtained by solving the Kohn-Sham equation,
\begin{equation}
  \left( -{\frac{\hbar^2}{2 m_e}} \nabla^2 
  + v_s^\sigma[\rho_\alpha,\rho_{\ensuremath{\mathsmaller\beta}}]({\bf r}) \right) \psi_{p\sigma}({\bf r}) = \epsilon_{p\sigma} \psi_{p\sigma}({\bf r}) \, .
  \label{eq:tddft.2}
\end{equation}
This involves a single-particle potential,
\begin{equation}
  v_s^\sigma[\rho_\alpha,\rho_{\ensuremath{\mathsmaller\beta}}]({\bf r}) = v_{ext}({\bf r}) +v_{\ensuremath{\mathsmaller{H}}}[\rho]({\bf r})
  +v_{xc}^\sigma[\rho_\alpha,\rho_{\ensuremath{\mathsmaller\beta}} ]({\bf r}) \, .
  \label{eq:tddft.3}
\end{equation}
which is the sum of the external potential $v_{ext}$ (i.e., the interaction of the electrons with the 
nuclei and any applied electronic fields), the Hartree potential,
\begin{equation}
  v_{\ensuremath{\mathsmaller{H}}}[\rho]({\bf r}) = \int \frac{\rho({\bf r}')}{\vert {\bf r}-{\bf r}' \vert} \, d{\bf r}' \, ,
  \label{eq:tddft.4}
\end{equation}
which may alternatively be obtained as the solution of Poisson's equation,
\begin{equation}
  \nabla^2 v_{\ensuremath{\mathsmaller{H}}}[\rho]({\bf r}) = -4 \pi \rho({\bf r}) \, ,
  \label{eq:tddft.5}
\end{equation}
and the xc-potential which is just the functional derivative of the xc-energy,
\begin{equation}
  v_{xc}^\sigma[\rho_\alpha,\rho_{\ensuremath{\mathsmaller\beta}}]({\bf r}) = \frac{\delta E_{xc}[\rho_\alpha,\rho_{\ensuremath{\mathsmaller\beta}}]}{\delta \rho_\sigma({\bf r})} \, .
  \label{eq:tddft.6}
\end{equation}
Together $v_{\ensuremath{\mathsmaller{H}}}[\rho]$ and $v_{xc}^\sigma[\rho_\alpha,\rho_{\ensuremath{\mathsmaller\beta}}]$ constitute the self-consistent field (SCF), 
$v_{{\ensuremath{\mathsmaller{H}}}xc}^\sigma[\rho_\alpha,\rho_{\ensuremath{\mathsmaller\beta}}]$.
Modern DFT often uses a type of generalized Kohn-Sham formalism where the SCF may contain some
orbital dependence due to (say) inclusion of some fraction of Hartree-Fock exchange and the external potential may
even include a nonlocal part.  In particular this latter generalization is the case when
nonlocal pseudopotentials are employed, as is the case in {\sc BigDFT}.

The external potential is time-dependent in TD-DFT, and the time-dependent Kohn-Sham potential has the
form,
\begin{equation}
  \left( -{\frac{\hbar^2}{2 m_e}} \nabla^2 
  + v_s^\sigma[\rho_\alpha,\rho_{\ensuremath{\mathsmaller\beta}}; \Phi_0;\Psi_0]({\bf r},t) \right) \psi_{p\sigma}({\bf r},t) 
  = i \hbar \frac{\partial }{\partial t} \psi_{p\sigma}({\bf r}) \, ,
  \label{eq:tddft.7}
\end{equation}
where the external potential,
\begin{equation}
  v_s^\sigma[\rho_\alpha,\rho_{\ensuremath{\mathsmaller\beta}};\Phi_0;\Psi_0]({\bf r},t) = v_{ext}({\bf r},t) +v_{\ensuremath{\mathsmaller{H}}}[\rho]({\bf r},t)
 +v_{xc}^\sigma[\rho_\alpha,\rho_{\ensuremath{\mathsmaller\beta}};\Phi_0;\Psi_0]({\bf r},t) \, .
  \label{eq:tddft.8}
\end{equation}
The dependence of the xc part on the initial interacting ($\Psi_0$) and noninteracting ($\Phi_0$) 
wave functions may be eliminated by using the first Hohenberg-Kohn theorem \cite{HK64} if the
initial state is the ground stationary state.  This is case when seeking the linear response of 
the ground stationary state to an applied electronic field.  The TD charge density,
\begin{equation}
  \rho({\bf r},t) = e \sum_{p\sigma} n_{p\sigma} \vert \psi_{p\sigma} ({\bf r},t)\vert^2 \, ,
  \label{eq:tddft.9}
\end{equation}
then suffices to calculate the induced dipole moment (at least for finite systems such as molecules)
and hence the dynamic polarizability,
\begin{equation}
   \alpha(\omega) = \sum_I^{I\neq 0} \frac{e^2 f_{\ensuremath{\mathsmaller{I}}}}{m_e(\omega_{\ensuremath{\mathsmaller{I}}}^2-\omega^2)} \, ,
   \label{eq:tddft.10}
\end{equation}
whose poles give the excitation energies of the system,
\begin{equation}
 \hbar \omega_{\ensuremath{\mathsmaller{I}}} = E_{\ensuremath{\mathsmaller{I}}}-E_0   \, ,
  \label{eq:tddft.10a}
\end{equation}
and whose residues give the corresponding oscillator strengths, 
\begin{equation}
  f_{\ensuremath{\mathsmaller{I}}} = \frac{2 \omega_{\ensuremath{\mathsmaller{I}}} m_e}{\hbar} \vert \langle \Psi_{\ensuremath{\mathsmaller{I}}} \vert {\bf r} \vert \Psi_0 \rangle \vert^2 \, .
  \label{eq:tddft.10b}
\end{equation}

The excitation energies and oscillator strengths are often presented in the form of a stick spectrum
consisting of lines of height $f_{\ensuremath{\mathsmaller{I}}}$ located at the associated $\omega_{\ensuremath{\mathsmaller{I}}}$.  What is actually measured
is the molar extinction coefficient $\epsilon$, in Beer's law.  To a first approximation, this
is related to the spectral function,
\begin{equation}
  S( \omega) = \sum_I f_{\ensuremath{\mathsmaller{I}}} \delta( \omega-\omega_{\ensuremath{\mathsmaller{I}}}) \, ,
  \label{eq:tddft.10c}
\end{equation}
by,
\begin{equation}
  \epsilon(\omega) = \frac{\pi N_{\ensuremath{\mathsmaller{A}}} e^2}{2 \epsilon_0 m_e c \ln (10)} S(\omega) \, ,
  \label{eq:tddft.10d}
\end{equation}
in SI units \cite{H82,H83,H02}.  Finite spectrometer resolution, vibrational, and solvent effects are traditionally approximated by replacing
the Dirac delta functions in Eq.~(\ref{eq:tddft.10c}) with a Gaussian whose full-width at half-maximum (FWHM)
is selected to match the experimental spectrum and we will do the same.  Solvent effects may also shift excitation energies
\cite{BM54a,BM54b,L57,LP57,M57,B64,BSK76} and affect oscillator strengths 
\cite{C34,MR41,W64,L66,BW68,A70,MB80,AI86,A91,HB94,BGK06}.
While these effects are not necessarily small ({\it vide infra}), we will simply ignore them 
when comparing with experimental data as this is a reasonable first approximation.

Practical TD-DFT calculations typically make use of the TD-DFT adiabatic approximation,
\begin{equation}
  v_{xc}^\sigma[\rho_\alpha,\rho_{\ensuremath{\mathsmaller\beta}}]({\bf r},t) = \frac{\delta E_{xc}[\rho^t_\alpha, \rho^t_{\ensuremath{\mathsmaller{\beta}}}]}
                      {\delta \rho^t_\sigma({\bf r})} \, ,
  \label{eq:tddft.11}
\end{equation}
because even less is known about the exact xc-potential in TD-DFT than is the case for 
regular DFT.  Here $\rho^t_\sigma({\bf r})$ is $\rho({\bf r},t)$ regarded at fixed $t$ as a function of 
${\bf r} =(x,y,z)$.  If in addition, the xc-energy functional is of the
local density approximation (LDA) or generalized gradient approximation (GGA) type, the TD-DFT adiabatic approximation is typically found to
work well for low-energy excitations of dominant one-hole/one-particle character which are not too
delocalized in space and do not involve too much charge transfer.  Several extensions of classic TD-DFT
have been found to be useful for going beyond these restrictions (see for example Ref.~\cite{C09,CH12} for a 
review).

One of us used a density-matrix formalism to develop  
a random-phase approximation (RPA) like formalism for the 
calculation of absorption spectra from the dynamic polarizability [Eq.~(\ref{eq:tddft.10})] \cite{C95}.
This allowed the calculation of TD-DFT spectra to be quickly implemented in a wide variety of quantum
chemistry programs since most of the available computational framework was already in place.  The precise
equation that needs to be solved is, within the TD-DFT adiabatic approximation,
\begin{equation}
  \left[  \begin{array}{cc} {\bf A} &{\bf  B} \\ {\bf B^*} & {\bf A^*} \end{array} \right ] 
  \left( \begin{array}{c} { \vec {X}} \\ { \vec{ Y} } \end{array}\right )
  = \hbar \omega \left[ \begin{array}{cc} {\bf 1} & {\bf 0} \\ {\bf 0} & -{\bf 1} \end{array} \right]
  \left( \begin{array}{c} {\vec { X}} \\ {\vec { Y} } \end{array}\right ) \, .
  \label{eq:tddft.12}
\end{equation}
Here,
\begin{eqnarray}
  A_{ai\sigma,bj\tau} & = & \delta_{ab} \delta_{ij}\delta_{\sigma\tau} (\epsilon_a -\epsilon_i) 
   + K_{ai\sigma,bj\tau} \nonumber \\
  B_{ai\sigma,bj\tau} & = & K_{ai\sigma,jb\tau} \, ,
  \label{eq:tddft.13}
\end{eqnarray}
and the coupling matrix is given by,
\begin{equation}
  K_{pq\sigma,rs\tau}   =   e^2 \int \int \psi_{p\sigma}^{*}({\bf r}) \psi_{q\sigma}({\bf r})  
   \left [ \frac{1}{\vert {\bf r} - {\bf r}' \vert} 
   + \frac {\delta^2 E_{xc}}{\delta \rho_\sigma({\bf r}) \delta \rho_\tau ({\bf r}')} \right ] 
  \psi_{r\tau}({\bf r}')\psi_{s\tau}^{*}({\bf r}') d{\bf r} d{\bf r}' \, ,
  \label{eq:tddft.14}
\end{equation}
where we are making use of the index convention that orbitals $a, b, \cdots, g,h$ are unoccupied,
orbitals $i,j,k,l,m,n$ are occupied, and orbitals $o,p,\cdots, x,y,z$ are free to be either 
occupied or unoccupied.  In the case of an LDA or GGA, Eq.~(\ref{eq:tddft.12}) may be rearranged to give 
the lower-dimensional matrix equation,
\begin{equation}
  {\bf \Omega}  {\vec F} = \hbar^2 \omega^2 {\vec { F}} \, ,
  \label{eq:tddft.15}
\end{equation}
where
\begin{equation}
  \Omega_{ia\sigma,jb\tau}  =  
  \delta_{ia}\delta_{jb}\delta_{\sigma\tau}{(\epsilon_{a\sigma} -\epsilon_{i\sigma})^2} + 
  2 \sqrt{(\epsilon_{a\sigma} -\epsilon_{i\sigma})} 
  K_{ia\sigma,jb\tau} \sqrt{(\epsilon_{a\sigma} -\epsilon_{i\sigma})} \, .
  \label{eq:tddft.16}
\end{equation}
Alternatively the Tamm-Dancoff approximation (TDA) is sometimes found to be useful \cite{HH99},
\begin{equation}
  {\bf A} {\vec { X}} = \hbar \omega {\vec { X}} \, .
  \label{eq:tddft.17}
\end{equation}
This is particularly the case when it is necessary to attenuate the effect of spin-instabilities on potential 
energy surfaces when investigating photoprocesses \cite{CND11}.


\section{Implementation in \sc{BigDFT\/}}
\label{sec:BigDFT}
%
Our implementation of the equations of the previous section in {\sc BigDFT} 
is described in this section and their validation
by comparison against calculations with {\sc deMon2k} is described in the next section. 

{\sc BigDFT} solves the Kohn-Sham equation [Eq.~(\ref{eq:tddft.2})] in the pseudopotential approximation.
The main difference is that the external potential part, $v_{ext}$, of the Kohn-Sham potential, $v_s^\sigma$,
is manipulated so as to smooth the behavior of the Kohn-Sham orbitals in the core region near
the nuclei while preserving the form of the Kohn-Sham orbitals outside the core region.  This is done through
the use of Goedecker-Teter-Hutter (GTH) pseudopotentials \cite{GTH96} in order to avoid ``wasting'' wavelets 
on describing the nuclear cusp.  These include both a local and nonlocal part and are used for all atoms 
(even hydrogen).  Several different functionals are available for the xc-energy. However we will only be
considering the LDA functional with Teter's Pade approximation \cite{GTH96} of Ceperley and Alder's quantum Monte Carlo results  \cite{CA80} since the
xc-kernel, $f_{xc}$, is thus far only implemented in {\sc BigDFT} at the LDA level.

\begin{figure}
\includegraphics[angle=0,width=0.6\textwidth]{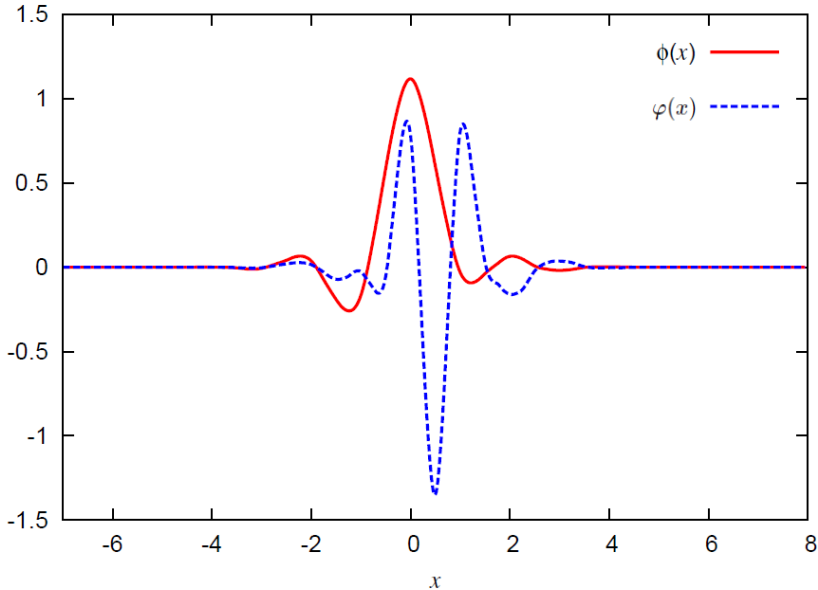}
\caption{Daubechies scaling function $\phi(x)$ and wavelet $\varphi(x)$ of order  16. \label{fig:daubechies} }
\end{figure}

There are two fundamental functions in Daubechies family: the
scaling function $\phi(x)$ and the wavelet $\varphi(x)$ (see
Fig. \ref{fig:daubechies}.) Note that both types of function
are localized with compact support. The full basis set can be
obtained from all translations by a certain grid spacing {\it h}
of the scaling and wavelet functions centered at the origin. 
These functions
satisfy the fundamental defining equations,
\begin{eqnarray}
\phi(x) & = & \sqrt{2} \sum_{j=1-m}^m h_j \phi(2x-j) \nonumber \, , \\
\varphi(x) & = & \sqrt{2} \sum_{j=1-m}^m g_j \phi(2x-j) \, .
\label{eq.bigdft1}
\end{eqnarray}
which relate the basis functions on a grid with spacing $ h$ and
another one with spacing $ h/2$.
The coefficients, h$_j$ and g$_j$, consitute the so-called ``filters"
which define the wavelet family of order $m$.  These coefficients
satisfy the relations, $\sum_j h_j = 1$ and $g_j = (-1)^j h_{-j+1}$. 
Equation (\ref{eq.bigdft1}) is very important since it means that
a scaling-function basis defined over a fine grid of spacing $h/2$
may be replaced by combining a scaling-function basis over a
coarse grid of spacing $h$ with a wavelet basis defined over
the fine grid of spacing $h/2$.  This then gives us the liberty
to begin with a coarse description in terms of scaling functions
and then add wavelets only where a more refined description is needed.
In principle the refined wavelet description may be further refined
by adding higher-order wavelets where needed.  However in {\sc BigDFT}
we restricted ourselves to just two levels: coarse and fine associated
respectively with scaling functions and wavelets.
 
For a three-dimensional description, the simplest basis set is
obtained by a tensor product of one-dimensional basis functions.  
For a two resolution level description, the
coarse degrees of freedom are expanded by a single three dimensional
function, $\phi_{i_1,i_2,i_3}^0(\bf r)$, while the fine degrees of freedom
can be expressed by adding  another seven basis functions, $\phi_{j_1,j_2,j_3}^{\nu}(\bf r)$,
which include tensor products with one-dimensional wavelet functions.
Thus, the Kohn-Sham wave function $\psi(\bf r)$ is of the form
\begin{equation}
\psi({\bf r}) = \sum_{i_1,i_2,i_3} c_{i_1,i_2,i_3}^0 \phi_{i_1,i_2,i_3}^0({\bf r}) +\sum_{j_1,j_2,j_3}\sum_{\nu_1}^7 c_{j_1,j_2,j_3}^{\nu} \phi_{j_1,j_2,j_3}^{\nu} ({\bf r}) \, .
\end{equation}
The sum over {\it{i$_1$,i$_2$,i$_3$}} runs over all the grid points contained
in the low-resolution regions and the sum over {\it{j$_1$,j$_2$,j$_3$}} runs over all
the points contained in the (generally smaller) high resolution regions. 
 Each wave function is then
described by a set of coefficients $\{c_{j_1,j_2,j_3}^{\nu}\},\nu=0,...,7$.  
Only the nonzero
scaling function and wavelet  coefficients are stored. The data is thus
compressed.  The basis set being orthogonal, several operations
such as scalar products among different orbitals and between orbitals
and the projectors of the nonlocal pseudopotential can be directly
carried out in this compressed form.
In addition to raw Daubechies scaling functions, practical
applications make use of autocorrelated functions to make
interpolating scaling functions (ISF) \cite{DD89}. In particular, as
shown by A. Neelov and
S. Goedecker \cite{NG+06}, the local potential matrix elements approximated
using the linear combination of such ISFs is
exact for polynomial expansions up to 7th order and the corresponding Kohn-Sham
densities can be calculated by the
real-valued coefficients on the grid points.

Although not every grid point is associated with a basis function and the fine grid is only used
in some regions of space, the Daubechies basis set is still very large.  This means that full
diagonalization of the Kohn-Sham orbital Hamiltonian is not possible.  Instead the  
direct minimisation method \cite{D75,MRD91} is used to obtain the  
occupied orbitals.  This is in contrast with the GTO-based {\sc deMon2k} program which will
be described below in which full diagonalization of the Kohn-Sham orbital 
Hamiltonian matrix is carried out within the
finite GTO basis set.

A key point to review because of its importance for our implementation of TD-DFT in {\sc BigDFT} is the
Poisson solver used to treat the Hartree part of the potential, $v_{\ensuremath{\mathsmaller{H}}}$.  
Although this Poisson solver has been discussed elsewhere \cite{GDN+06,GDG07}, we briefly review 
how it works in order to keep this article somewhat self-contained.
The Hartree potential is evaluated as,
\begin{equation}
  v_{\ensuremath{\mathsmaller{H}}}({\bf r}) = \int G(\vert {\bf r}-{\bf r}' \vert) \rho({\bf r'}) \, d{\bf r}' \, ,
  \label{eq:BigDFT.1}
\end{equation}
where $G(r) = 1/r$ is the Green's function for the Poisson equation, namely just the Coulomb
potential.  The density and potential are expanded in a set of interpolating scaling functions,
\begin{eqnarray}
  \rho({\bf r}) & = & \sum_{i_1,i_2,i_3}\tilde \phi_{i_1}(x) \tilde \phi_{i_2}(y) \tilde \phi_{i_3}(z)
                      \rho_{i_1,i_2,i_3}
   \nonumber \\
  v_{\ensuremath{\mathsmaller{H}}}({\bf r}) & = & \sum_{i_1,i_2,i_3} \tilde \phi_{i_1}(x) \tilde \phi_{i_2}(y) \tilde \phi_{i_3}(z)
                      v_{i_1,i_2,i_3} \, ,
  \label{eq:BigDFT.2}
\end{eqnarray}
associated with the same grid of points, ${\bf r}_{i_1,i_2,i_3}$, in real space.  In particular, 
the charge density coefficients, $\rho_{i_1,i_2,i_3} = \rho({\bf r}_{i_1,i_2,i_3})$.  Then,
\begin{equation}
  v_{i_1,i_2,i_3} = \sum_{j_1,j_2,j_3} G_{j_1-i_1,j_2-i_2,j_3-i_3} \rho_{j_1,j_2,j_3}
  \, ,
  \label{eq:BigDFT.3}
\end{equation}
where the quantity,
\begin{equation}
  G_{j_1-i_1,j_2-i_2,j_3-i_3} = \int \frac{\tilde \phi_{j_1}(x') \tilde \phi_{j_2}(y') \tilde \phi_{j_3}(z')}
                                {\vert {\bf r}_{i_1,i_2,i_3} - {\bf r}' \vert} \, d{\bf r}'
  \, ,
  \label{eq:BigDFT.4}
\end{equation}
is translationally invariant by construction.  Since Eq.~(\ref{eq:BigDFT.3}) has the form of a three
dimensional convolution, it may be efficiently evaluated by using appropriate parallelized fast
Fourier transform algorithms at the cost of only ${\cal O}(N \ln N)$ operations.
The calculation of matrix elements of the Green's function $G(r) = 1/r$ is simplified by
using a separable approximation in terms of Gaussians,
\begin{equation}
  \frac{1}{r} \approx \sum_{k} e^{-p_k r^2} c_k \, ,
  \label{eq:BigDFT.5}
\end{equation}
so that all the complicated 3-dimensional integrals are reduced to products of simpler 1-dimensional
integrals.
For more information about {\sc BigDFT}, the
reader is referred to the program website \cite{bigdft} and to various publications
\cite{GDN+06,GDG07,GNG+08,AGGNG09,GODMNG09,GVO+11}.

We are now in a position to understand the construction of the coupling matrix in our
implementation of TD-DFT in {\sc BigDFT}, which we split into the Hartree and exchange-correlation
parts,
\begin{equation}
  K_{ai\sigma,bj\tau} = K^{{\ensuremath{\mathsmaller{H}}}}_{ai\sigma,bj\tau} + K^{xc}_{aj\sigma,bj\tau} \, .
  \label{eq:BigDFT.6}
\end{equation}
Instead of calculating the Hartree part of coupling matrix directly as,
\begin{equation}
  K^{{\ensuremath{\mathsmaller{H}}}}_{ai\sigma,bj\tau}  = \int \int \psi_{a\sigma}^{*}({\bf r}) \psi_{i\sigma}({\bf r})
         \frac{1}{\vert {\bf r}  - {\bf r}' \vert}
         \psi_{b\tau}({\bf r'}) \psi_{j\tau}^{*}({\bf r'}) \, d{\bf r} d{\bf r}' \, ,
  \label{eq:BigDFT.7}
\end{equation}
we express the coupling matrix element as,
\begin{equation}
  K^{{\ensuremath{\mathsmaller{H}}}}_{ai\sigma,bj\tau}  = \int \psi_{a\sigma}^{*}({\bf r}) \psi_{i\sigma}({\bf r})
                v_{bj\tau}({\bf r}) \, d{\bf r} \, ,
  \label{eq:BigDFT.8}
\end{equation}
where,
\begin{equation}
  v_{ai\sigma}({\bf r}) = \int \frac{\rho_{ai\sigma}({\bf r})}{\vert {\bf r} - {\bf r}'|} \, d{\bf r}' \, ,
  \label{eq:BigDFT.9}
\end{equation}
and,
\begin{equation}
  \rho_{ai\sigma}({\bf r}) = \psi_{a\sigma}^{*}({\bf r})\psi_{i\sigma}({\bf r}) \, .
  \label{eq:BigDFT.10}
\end{equation}
The advantage of doing this is that, although $\rho_{ai\sigma}$ and $v_{ai\sigma}$ 
are neither real physical charge densities nor real physical potentials, they still 
satisfy the Poisson equation,
\begin{equation}
  \nabla^2 v_{ai\sigma}({\bf r}) = - 4 \pi \rho_{ai\sigma}({\bf r}) \, ,
  \label{eq:BigDFT.11}
\end{equation}
and we can make use of whichever of the efficient wavelet-based Poisson solvers
already available in {\sc BigDFT}, is appropriate for the boundary conditions of our
physical problem.  

Once the solution of Poisson's equation, $v_{ai\sigma}({\bf r})$, is known, we can then calculate the
Hartree part of the kernel according to Eq.~(\ref{eq:BigDFT.8}).  Inclusion of the exchange-correlation
kernel is accomplished by evaluating,
\begin{equation}
  K_{ai\sigma,bj\tau} = \int M_{ai\sigma}({\bf r}) \rho_{bj\tau}({\bf r}) \, d{\bf r}
  \, ,
  \label{eq:BigDFT.12}
\end{equation}
where,
\begin{equation}
  M_{ai\sigma}({\bf r}) = v_{ai\sigma}({\bf r}) 
   +  \int \rho_{ai\sigma} ({\bf r}')  f_{xc}^{\sigma,\tau}({\bf r},{\bf r}')  \, d{\bf r}'
  \, .
  \label{eq:BigDFT.13}
\end{equation}
We note that $f_{xc}^{\sigma,\tau}({\bf r},{\bf r}') = f_{xc}^{\sigma,\tau}({\bf r},{\bf r}') \delta({\bf r}-{\bf r}')$ for the LDA, so that no integral need actually be carried out in evaluating $M_{ai\sigma}({\bf r})$.
The integral in Eq.~(\ref{eq:BigDFT.12}) is, of course, carried out 
numerically in practice as a discrete. 


\section{Validation}
\label{sec:validation}
%
Having implemented TD-DFT in {\sc BigDFT} we now desire to validate our implementation by testing it against
another program in which TD-DFT is already implemented, namely the all-electron GTO-based program {\sc deMon2k} \cite{deMon2k}.
{\sc deMon2k} resembles a typical GTO-based quantum chemistry program in that all the integrals 
other than the xc-integrals, can be evaluated analytically.  
In particular, {\sc deMon2k} has the important advantage that it accepts the popular GTO basis sets common in quantum chemistry
and so can benefit from the experience in basis set construction of a large community built up over the past 50 years or so.
In the following, we have chosen to use the well-known correlation-consistent basis sets for this study \cite{F96,SDESGCW07}.
(Note, however, that the correlation-consistent basis sets used in {\sc deMon2k}
lack $f$ and $g$ functions but are otherwise exactly the same as the usual ones.) 
The advantage of using these particular basis sets is that there is a clear 
hierarchy as to quality.

An exception to the rule that integrals are evaluated analytically in {\sc deMon2k} are the xc-integrals (for the xc-energy, xc-potential, and
xc-kernel) which are evaluated numerically over a Becke atom-centered grid.  This is important because the relative
simplicity of evaluating integrals over a grid has allowed the rapid implemenation of new functionals as they were introduced.
We made use of the fine fixed grid in our calculations.  

As described so far, {\sc deMon2k} should have ${\cal O}(N^4)$ scaling because of the need to evaluate 4-center integrals.
Instead {\sc deMon2k} uses a second atom-centered auxiliary GTO basis to expand the charge density.  This allows the
the elimination of all 4-center integrals so that only 3-center integrals remain for a formal ${\cal O}(N^3)$ scaling.
In practice, integral prescreening leads to ${\cal O}(N^M)$ scaling where $M$ is typically between 2 and 3.
We made use of the A3 auxiliary basis set from the {\sc deMon2k} automated auxiliary basis set library.

All calculations were performed using standard {\sc deMon2k} default criteria.
Although full TD-LDA calculations are possible with {\sc deMon2k}, the TD-LDA calculations reported here all made use
of the TDA.
The chosen test molecule was N$_2$ with an optimized bond length of 1.115 \AA.
This molecule was chosen partly because of its small size but also because
of the large number of excited states which are well characterized (see Refs. \cite{JCS96,CJC+98} and references therein.)

Unlike TD Hartree-Fock (or configuration interaction singles) calculations, TD-LDA calculations
are preprepared to describe excitation
processes in the sense that the occupied and unoccupied orbitals both see the same number of electrons
(because they come from the same local potential).  This means that there is often little
relaxation --- at least in small molecules --- and a two orbital model \cite{C95,C08,CH12,CGG+00} often provides a good first
approximation to the singlet ($\hbar \omega^{\ensuremath{\mathsmaller{S}}}_{i \rightarrow a}$) and triplet
($\hbar \omega^{\ensuremath{\mathsmaller{T}}}_{i \rightarrow a}$) excitation energies,
\begin{eqnarray}
  \hbar \omega^{\ensuremath{\mathsmaller{T}}}_{i \rightarrow a} & = & \epsilon_a - \epsilon_i
     + (ia \vert f_{xc}^{\alpha,\alpha}-f_{xc}^{\alpha,\beta} \vert ai) \nonumber \\
  \hbar \omega^{\ensuremath{\mathsmaller{S}}}_{i \rightarrow a} & = & \epsilon_a - \epsilon_i
     + (ia \vert 2f_{\ensuremath{\mathsmaller{H}}} +f_{xc}^{\alpha,\alpha}+f_{xc}^{\alpha,\beta} \vert ai) \, .
  \label{eq:BigDFT.14}
\end{eqnarray}
Consideration of typical integral signs and magnitudes then implies that
\begin{equation}
  \hbar \omega^{\ensuremath{\mathsmaller{T}}}_{i \rightarrow a} \leq \epsilon_a - \epsilon_i \leq \hbar \omega^{\ensuremath{\mathsmaller{S}}}_{i \rightarrow a} \, ,
  \label{eq:BigDFT.15}
\end{equation}
with the singlet-triplet splitting going to zero for Rydberg states (in which case the electron
repulsion integrals become negligible due to the diffuse nature  of the target orbital $\psi_a$.)

\begin{table}
  \label{basis-set-dependence}
  \caption{Basis set dependence of the HOMO and LUMO energies and of the HOMO-LUMO gap (eV) calculated
           using {\sc deMon2k}.  \label{tab:deMon-basis-set-dependence}}
  \begin{tabular}{cccc}
  \hline \hline
  Basis Set & $-\epsilon_{\ensuremath{\mathsmaller{{\text{HOMO}}}}}$ & $-\epsilon_{\ensuremath{\mathsmaller{{\text{LUMO}}}}}$
  &  $\Delta \epsilon_{\ensuremath{\mathsmaller{{\text{HOMO}-\text{LUMO}}}}}$  \\
  \hline
  STO-3G     &  7.6758  & 0.0297      &   7.6461   \\
  \multicolumn{4}{c}{ } \\
  DZVP       &  10.1824 & 2.2616      &   7.9208  \\
  TZVP       &  10.2142 & 2.2894      &   7.9248  \\
  \multicolumn{4}{c}{ } \\
  CC-PVDZ        &  9.8656  & 1.8993      &   7.9663  \\
  CC-PVTZ        &  10.2978 & 2.2868      &   8.0110  \\
  CC-PVQZ        &  10.3545 & 2.3527      &   8.0018 \\
  CC-PV5Z        &  10.3786 & 2.3886      &   7.9900  \\
 \multicolumn{4}{c}{ } \\
  CC-PCVDZ       &  9.9197  & 1.9314      &   7.9883  \\
  CC-PCVQZ       &  10.3555 & 2.3532      &   8.0023  \\
  CC-PCVTZ       &  10.2372 & 2.2718      &   8.0154  \\
  CC-PCV5Z       &  10.3793 & 2.3891      &   7.9902  \\
  \multicolumn{4}{c}{ } \\
  AUG-CC-PVDZ    &  10.3534 & 2.3785      &   7.9749  \\
  AUG-CC-PVQZ    &  10.3987 & 2.4127      &   7.9860 \\
  AUG-CC-PVTZ    &  10.3953 & 2.4010      &   7.9943  \\
  AUG-CC-PV5Z    &  10.3984 & 2.4137      &   7.9847  \\
  \multicolumn{4}{c}{ } \\
  AUG-CC-PCVDZ   &  10.3732 & 2.3879      &   7.9847  \\
  AUG-CC-PCVTZ   &  10.3972 & 2.4015      &   7.9957  \\
  AUG-CC-PCVQZ   &  10.3990 & 2.4124      &   7.9866  \\
  AUG-CC-PCV5Z   &  10.3985 & 2.4136      &   7.9849  \\
  \hline \hline
  \end{tabular}
\end{table}

\begin{table}
  \caption{Basis set dependence of the HOMO and LUMO energies and of the HOMO-LUMO gap (eV) calculated
           using {\sc BigDFT}.  \label{tab:BigDFT-basis-set-dependence}}
  \begin{tabular}{cccc}
  \hline
  h$_g$\footnotemark[1]/m\footnotemark[2]/n\footnotemark[3]
  & $-\epsilon_{\ensuremath{\mathsmaller{{\text{HOMO}}}}}$ & $-\epsilon_{\ensuremath{\mathsmaller{{\text{LUMO}}}}}$
  &  $\Delta \epsilon_{\ensuremath{\mathsmaller{{\text{HOMO}-\text{LUMO}}}}}$  \\
  \hline
  \multicolumn{4}{c}{Low resolution}\\
  0.4/6/8  & 10.3910   & 2.3815  & 8.0095 \\
  0.4/7/8  &  10.3964  & 2.3922  & 8.0042 \\
  0.4/8/8  & 10.3971   & 2.3945  & 8.0027 \\
  0.4/9/8  & 10.3972   & 2.3951  & 8.0022 \\
  0.4/10/8 & 10.3973   & 2.3953  & 8.0020 \\
  0.4/11/8 & 10.3972   & 2.3953  & 8.0019 \\
  \multicolumn{4}{c}{High resolution}\\
  0.3/7/8  & 10.3977   & 2.3932  & 8.0043 \\
  0.3/8/8  & 10.3984   & 2.3957  & 8.0027 \\
  0.3/9/8  & 10.3985   & 2.3963  & 8.0022 \\
  0.3/10/8 & 10.3985   & 2.3965  & 8.0021 \\
  0.3/11/8 & 10.3985   & 2.3965  & 8.0020 \\
  \hline
  \end{tabular}
  \footnotetext[1]{ Grid spacing of the cartesian grid in atomic units.}
  \footnotetext[2]{ Coarse grid multiplier (crmult) }
  \footnotetext[3]{Fine grid multiplier (frmult)}
\end{table}
Since orbital energy differences provide an important first estimate of TD-DFT excitation energies,
we wished to see how rapidly they converged for {\sc BigDFT} and {\sc deMon2k} as the quality of
the basis set was improved.  Tables~\ref{tab:deMon-basis-set-dependence}
and \ref{tab:BigDFT-basis-set-dependence} show how the highest occupied
molecular orbital (HOMO) -- lowest unoccupied molecular orbital (LUMO) energy gap ($\Delta \epsilon_{\ensuremath{\mathsmaller{{\text{HOMO}-\text{LUMO}}}}}$)
varies for each program.

Consider first how {\sc deMon2k} calculations of $\Delta \epsilon_{\ensuremath{\mathsmaller{{\text{HOMO}-\text{LUMO}}}}}$
evolve as the basis set is improved (Table~\ref{tab:deMon-basis-set-dependence}).
Jamorski, Casida, and Salahub had earlier shown that LUMO is bound for reasonable basis sets \cite{JCS96}.
(Small differences between the present calculations and those in Ref.~\cite{JCS96} are due to gradual
improvements in the grid, auxiliary basis sets, and convergence criteria used in the {\sc deMon} programs).
Convergence to the true HOMO-LUMO LDA gap is expected  
with systematic improvement within the series: (i) double zeta plus valence polarization (DZVP)
$\rightarrow$ triple zeta plus valence polarization (TZVP),
(ii) augmented correlation-consistent double zeta plus polarization plus diffuse on all atoms (AUG-CC-PCVDZ) $\rightarrow$
AUG-CC-PCVTZ (triple zeta) $\rightarrow$ AUG-CC-PCVQZ (quadruple zeta) $\rightarrow$ AUG-CC-PCV5Z (quintuple zeta),
(iii) augmented correlation-consistent valence double zeta plus polarization
 plus diffuse (AUG-CC-PVDZ) $\rightarrow$
AUG-CC-PVTZ $\rightarrow$ AUG-CC-PVQZ $\rightarrow$ AUG-CC-PV5Z,
(iv) correlation-consistent double zeta plus polarization plus tight core 
(CC-PCVDZ) $\rightarrow$ CC-PCVTZ $\rightarrow$ CC-PCVQZ $\rightarrow$ CC-PCV5Z, and (v)
correlation-consistent valence double zeta plus polarization on all atoms (CC-PVDZ) $\rightarrow$ CC-PVTZ $\rightarrow$ CC-PVQZ 
$\rightarrow$ CC-PV5Z.
There is a clear tendency in the correlation-consistent basis sets to tend towards values of
10.40 eV for the HOMO energy, 2.42 eV for the LUMO energy, and 8.01 eV for
$\Delta \epsilon_{\ensuremath{\mathsmaller{{\text{HOMO}-\text{LUMO}}}}}$,
with adequate convergance already achieved with the CC-PVTZ basis set.

Now let us turn to {\sc BigDFT} (Table~\ref{tab:BigDFT-basis-set-dependence}).  Calculations were done for two different grid values, 
denoted by h$_g$ = 0.3, 0.4.
(These values are the nodes of the grid in atomic units which serve as centers
for the scaling function/wavelet basis.)
The simulation ``box" has the shape of the molecule and its size is
expressed in the units of  the coarse grid multiplier (crmult) and the fine grid multiplier (frmult)
which determines the radius for the
low/high resolution sphere around the atom.
Results using 16 unoccopied orbitals with the different wavelet basis sets are essentially identical,
with no significant variation in the HOMO energy, the LUMO energy, and
the  $\Delta \epsilon_{\ensuremath{\mathsmaller{{\text{HOMO}-\text{LUMO}}}}}$ value of 8.00 eV between
the high resolution combination of 0.3/11/8 and the low resolution combination of 0.4/6/8.

The remaining differences for the HOMO energy, LUMO energy, and  $\Delta \epsilon_{\ensuremath{\mathsmaller{{\text{HOMO}-\text{LUMO}}}}}$
calculated by the two programs, {\sc deMon2k} and {\sc BigDFT},  is more difficult to trace. For example,
it might
be due to the auxiliary basis approximation in {\sc deMon2k} or to the use of pseudopotentials in {\sc BigDFT}
or perhaps to still other program differences.  The important point is that differences are remarkably small.

\begin{figure}
\includegraphics[angle=0,width=0.8\textwidth]{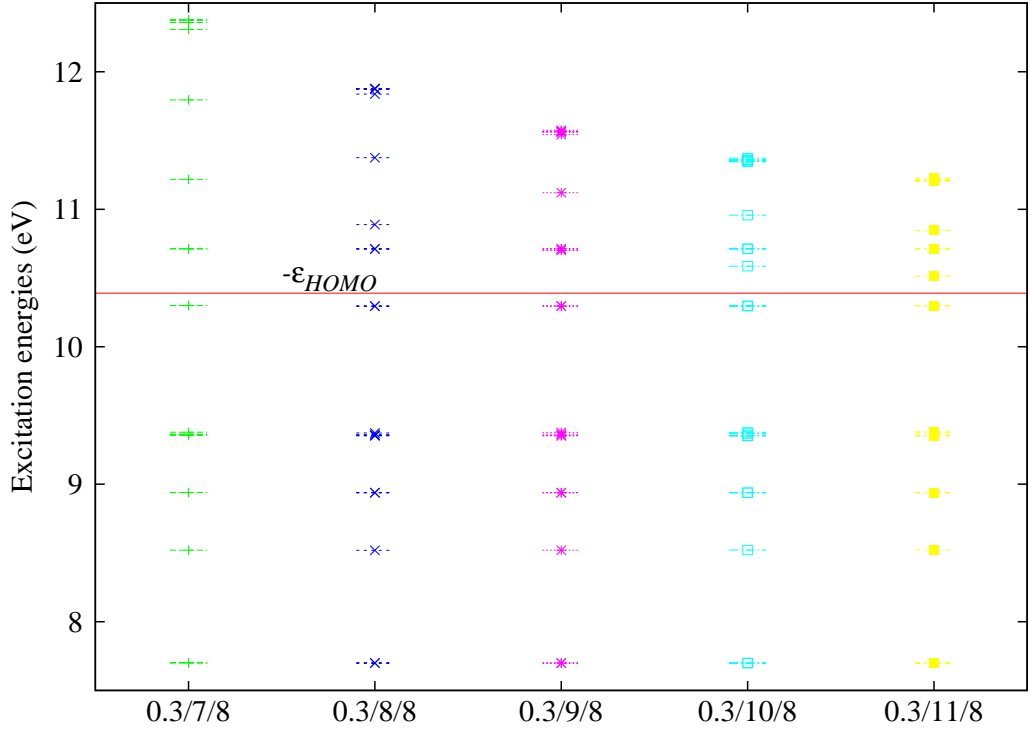}
\caption{ Singlet  and triplet excitation energies for N$_2$ 
         calculated using {\sc BigDFT}
          \label{fig:excited-states-continuum}
}
\end{figure}
\begin{table}
  \caption{Comparison of the nine lowest excitation energies of N$_2$ (in eV) calculated using
           different programs and with experiment. \label{tab:n2-excitation-energy}}
  \begin{tabular}{ccccc}
  \hline \hline
  State             & {\sc BigDFT}\tablenotemark[1]
                    &  {\sc deMon2k}\tablenotemark[2]
                    & {\sc ABINIT}\tablenotemark[3] & Experiment\tablenotemark[4]\\
  \hline
  1$^3\Sigma_g^{+}$  & 10.71         & 10.39           & 9.16         & 12.0 \\
  1$^3\Pi_u$         & 10.58         & 10.38           & 10.79        & 11.19\\
  1$^1\Delta_u$      & 10.29         & 9.99            & 10.46        & 10.27\\
  1$^1\Sigma_u^{-}$  & 9.37          & 9.36            & 9.92         & 9.92\\
  1$^3\Sigma_u^{-}$  & 9.36          & 9.36            & 9.91         & 9.67 \\
  1$^1\Pi_g$         & 9.35          & 9.10            & 9.47         & 9.31\\
  1$^3\Delta_u$      & 8.93          & 8.60            & 9.08         & 8.88 \\
  1$^3\Pi_g$         & 7.69          & 7.83            & 7.85         & 7.75 \\
  1$^3\Sigma_u^{+}$  & 8.52          & 8.43            & 8.16         & 8.04 \\
  \hline \hline
  \end{tabular}
  \tablenotetext[1]{Present work (TD-LDA/TDA).}  
  \tablenotetext[2]{Present work (TD-LDA/TDA).}
  \tablenotetext[3]{From \cite{abinit-tddft} (TD-LDA).}
  \tablenotetext[4]{Taken from Ref.(\cite{BK90}).}
\end{table}

We now come to the calculation of the actual excited states of N$_2$ and the third reason
alluded to in the introduction that quantum chemists have been slow to adapt grid-based
methods.  This is the concern that the very large size of basis sets in grid-based methods
would lead to impractically-large configuration interaction expansions.  Put differently,
this concerns the basic problem of how to handle the continuum. A correct inclusion of
the continuum in the formalism of Sec.~\ref{sec:TDDFT} would seem to require at least
approximate integrals over a quasicontinuum of unoccupied orbitals.  Clearly this is
impractical and our method only uses the first several unoccupied orbitals.  This is
perhaps reasonable for the lower excited states, given the anticipated dominance of the
two-orbital model [Eq.~(\ref{eq:BigDFT.14})], but may fail to be quantitative when relaxation
or state mixing starts to become important and the two-orbital model breaks down.  This
problem is avoided in quantum chemistry where the virtual orbitals in the TD-DFT calculation
have an altogether different meaning: they are simply there to describe the dynamic
polarizability of the ground-state charge density and need not describe the continuum
well.  This is why fewer unoccupied orbitals are normally needed in quantum chemical
applications of TD-DFT than is the case when using, say, plane waves.
In Fig.~\ref{fig:excited-states-continuum} the lowest few excited-states are calculated 
using {\sc BigDFT}.  
This figure shows that the
excited states calculated with different grids are essentially the same up to -$\epsilon_{\ensuremath{\mathsmaller{{\text{HOMO}}}}}$.  After -$\epsilon_{\ensuremath{\mathsmaller{{\text{HOMO}}}}}$, increasing
the quality of the basis set by varying the simulation box sizes leads to an increasing
collapse of the higher excited-states.
This is a reflection of the fact that the TD-LDA
ionization continuum starts at -$\epsilon_{\ensuremath{\mathsmaller{{\text{HOMO}}}}}$ which is 
about 5 eV low because of
the incorrect asymptotic behavior of the xc-potential \cite{CJC+98}.
Of course, as indicated by the estimate (\ref{eq:BigDFT.14}), the excitation energy is not exactly
a simple orbital energy difference, but they are closely related.  

With these caveats, let us see how well our implementation of TD-DFT does in {\sc BigDFT}.
Table~\ref{tab:n2-excitation-energy}, lists the lowest nine excited states of N$_2$ calculated with
{\sc deMon2k} and {\sc BigDFT}.  
For comparison, Table {\ref{tab:n2-excitation-energy}} also contains full TD-LDA (i.e, non-TDA) excitation
energies obtained from {\sc ABINIT} \cite{abinit} using the Perdew-Wang 92 parameterisation of the LDA 
functional \cite{PW92} along with the experimental values 
from the literature \cite{BK90}.
The slight differences which occur between the 
 1$^1\Sigma_u^{-}$ and  1$^3\Sigma_u^{-}$ excitation energies in the
{\sc BigDFT} and {\sc ABINIT} calculations are an indication
of residual numerical errors since these two states are
rigorously degenerate by symmetry when using the TD-LDA
and TD-LDA/TDA approximations: Aside from this tiny
difference, it is certainly reassuring that excitation energies calculated with {\sc BigDFT}, {\sc deMon2k},
and {\sc ABINIT} are quite similar. Nevertheless differences as large as 0.3 eV or more are found for some states.
Such differences are large enough to be potentially problematic for determining the ordering of near-lying
states.  
\begin{figure}
\includegraphics[angle=0,width=0.8\textwidth]{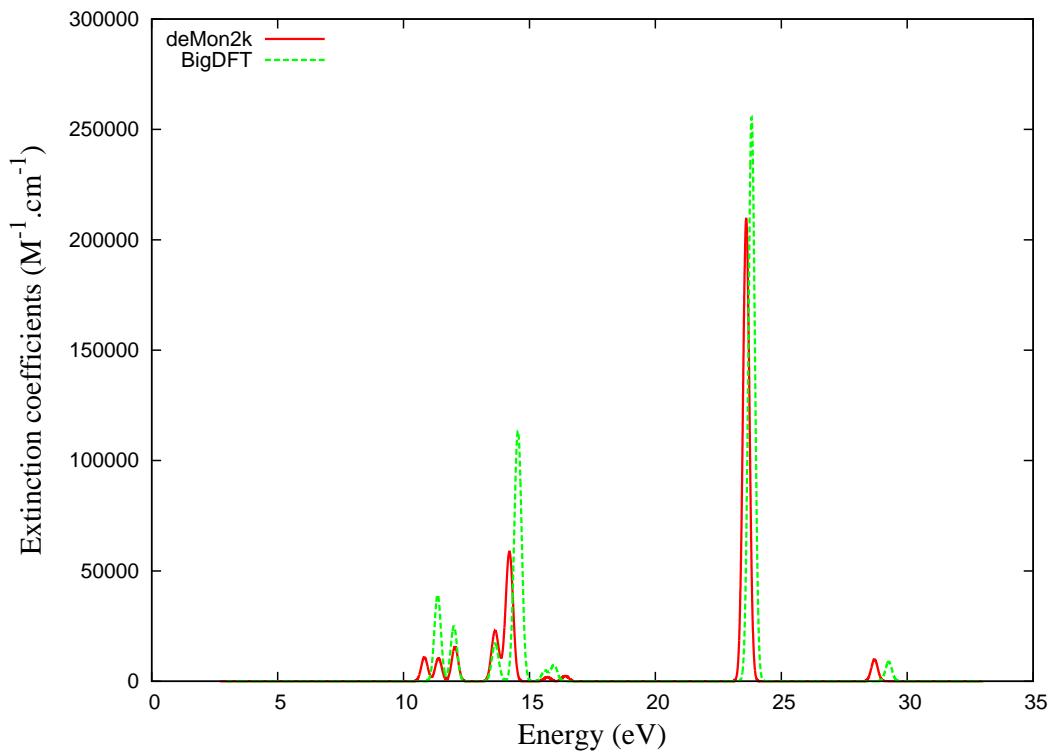}
\caption{Comparison of {\sc deMon2k} and {\sc BigDFT} N$_2$ spectra at higher energies.
         \label{fig:n2-spectra-bigdft-demon-0}
}
\end{figure}

Finally since a large number of excited states have been calculated, it is interesting to test the assertion that the oscillator strength
distribution should be approximately correct even above the TD-LDA ionization threshold at
-$\epsilon_{\ensuremath{\mathsmaller{{\text{HOMO}}}}}$ \cite{CJC+98}.  This is especially true since the transitions given in
Table~\ref{tab:n2-excitation-energy} are all dark and we would like to see how the oscillator
strengths of bright states compare.
The high-energy spectra are compared in Fig.~\ref{fig:n2-spectra-bigdft-demon-0}
using Eq.~(\ref{eq:tddft.10d}). 
Clearly the spectra are in
reasonable qualitative agreement.

The above results show that this first implementation of {\sc BigDFT} is quantitative --- especially
when results are dominated by bound-bound transitions ---  and
our discussion may eventually suggest ways to go further towards improving the method.  In the
meantime, we have a method that can be used for moderate-size molecules and this is illustrated in
the next section.

\section{Application}
\label{sec:flugi6}

A large series of fluorescent molecules of potential interest as biological markers \cite{GR03} has recently 
been synthesized  by combinatorial chemistry \cite{BMO11}.  
This is a method whereby large sets of similar reactions are conducted in parallel in arrays
of spots on a single plate, thus dramatically increasing throughput when searching for molecules with a
particular property --- in this case fluorescence.  
We have chosen to calculate the absorption spectrum of one of these fluorescent molecules
in preparation for future more in-depth theoretical studies of 
their fluorescence properties using {\sc BigDFT}.  This molecule, which we will simply refer to as Flugi {\bf 6} 
(because it is molecule 6 \cite{BMO11} among the fluorescent molecules prepared by the UGI reaction \cite{u62})
rather than by its full name of $N$-cyclohexyl-2-(4-methoxyphenyl)imidazo[1,2-$a$]pyridin-3-amine is shown 
in Fig.~\ref{fig:Olga_Flugi2}. The synthesis and partial characterization of Flugi {\bf 6} has been
described in Ref. \cite{BMO11}. However we go further here and report the experimental 
determination of its crystal structure. 
We then go on to compare the spectrum calculated with {\sc BigDFT} 
with the measured spectrum and discuss the problem of peak assignment.

\subsection{X-ray Crystal Structure}
Crystals of Flugi {\bf 6} (C$_{20}$H$_{23}$N$_3$O, $M$ = 321.42 g/mol) were obtained out of recrystallisation in 
 ethyl acetate (EtOAc) as colorless needles suitable for X-ray diffraction.  A 0.38 mm $\times$ 0.28 mm $\times$ 0.01 mm crystal
was mounted on a glass fiber using grease and centered on a Bruker Enraf Nonius kappa charge-coupled device (CCD) detector 
working at 200 K and at the monochromated (graphite) Mo K$_\alpha$ radiation $\lambda$  = 0.71073 {\AA}. The crystal was found to be 
orthorhombic, Pna2$_1$, $a$ = 27.912(4) {\AA} , $b$ = 5.876(2){\AA}, $c$ = 10.297(2) {\AA}, $V$ = 1688.7(6) {\AA}$^3$, $Z$ = 4, 
$D_x$ = 1.264 g.cm$^{-3}$, $m$ = 0.080 mm$^{-1}$. A total of 17700 reflections were collected using $\phi$ and $\omega$ scans; 
2853 independent reflections ($R_{int}$ = 0.1557). The data were corrected for the Lorentz and polarization effects. 
The structure was solved by direct methods with {\sc SIR92} \cite{ACGG93} and refined against F by leastsquare 
method implemented by TeXsan \cite{TeXsan}. C, N, and O atoms were refined anisotropically by the full matrix 
least-squares method. H atoms were set geometrically and recalculated before the last refinement cycle. 
The final $R$ values for 1964 reflections with  $I>2\sigma$ (I) and 217 parameters are $R1$ = 0.0617, $wR2$ = 0.0657, 
goodness of fit (GOF) = 1.78 and for all 2854 unique reflections $R1$ = 0.0923 , $wR2$ = 0.0829, GOF = 1.85. The resultant crystal geometry is given
in {\ref{tab:Flugi_6}}

\begin{table} [!h]
  \caption{Experimental geomentry (Cartesian coordinates in {\AA}) for the Flugi {\bf 6}  \label{tab:Flugi_6}}
  \begin{tabular}{cccc}
  \hline \hline
  Atom &   $x$ & $y$  &  $z$ \\
  \hline
O   &     14.0340&    1.5882&   2.9552  \\
N   &    7.7362  &  0.5932  &  7.4268\\
N   &    9.1303  &  2.5361  &  7.6402\\
N   &     8.7184 &  -0.6003 &  5.8202\\
C   &    8.8581  &  1.3112  &  7.0430\\
C   &    7.6978  & -0.5625  &  6.6613\\
C   &    6.7910  &  0.8587  &  8.3836\\
C   &    5.8275  & -0.0445  &  8.6210\\
C   &    9.4459  &  0.5452  &  6.0467\\
C   &   12.4814  &  2.2280  &  4.5690\\
C   &    9.6943  &  2.5959  &  8.9709\\
C   &   12.2450  &  0.0746  &  3.5887\\
C   &   12.9311  &  1.2651  &  3.6923\\
C   &   11.3663  &  2.0047  &  5.3306\\
C   &   11.1149  & -0.1284  &  4.3608\\
C   &   10.6426  &  0.8232  &  5.2590\\
C   &    5.7504  & -1.2331  &  7.8594\\
C   &    6.6706  & -1.4857  &  6.8881\\
C   &   11.0628  &  2.0293  &  9.0805\\
C   &   14.5914  &  0.5796  &  2.1251\\
C   &    9.6456  &  4.0228  &  9.4571\\
C   &   10.1882  &  4.1778  & 10.8473\\
C   &   11.6027  &  2.1400  & 10.5019\\
C   &   11.5598  &  3.5428  & 10.9936\\
H   &      6.8178&    1.6727&    8.8752\\
H   &      5.1910&    0.1163&    9.3101\\
H   &     12.9482&    3.0519&    4.6478\\
H   &      9.1307&    2.0846&    9.5420\\
H   &     12.5454&   -0.6011&    2.9948\\
H   &     11.0776&    2.6830&    5.9322\\
H   &     10.6436&   -0.9486&    4.2766\\
H   &      5.0541&   -1.8576&    8.0282\\
H   &      6.6161&   -2.2802&    6.3725\\
H   &   11.0384  &  1.1132  &  8.8327 \\
H   &   11.6406  &  2.5025  &  8.4958 \\
H   &    10.1595 &   4.5627 &   8.8705\\
H   &     8.7422 &   4.3165 &   9.4511\\
H   &    10.2551 &   5.1036 &  11.0519\\
H   &     9.5907 &   3.7627 &  11.4607\\
H   &    11.0761 &   1.5956 &  11.0779\\
H   &    12.5025 &   1.8356 &  10.5174\\
H   &    12.1865 &   4.0570 &  10.5007\\
H   &    11.8007 &   3.5496 &  11.9149\\
H   &    15.3192 &   0.9409 &   1.6353\\
H   &    14.8996 &  -0.1395 &   2.6665\\
H   &    13.9283 &   0.2626 &   1.5256\\
H   &     8.9445 &   3.3408 &   7.1731\\
 \hline  \hline
  \end{tabular}
\end{table}

The data have been deposited with the Cambridge Crystallographic Data Centre (Reference No. CCDC 8200007). 
This material is available free of charge via www.ccdc.can.ac.uk/conts/retrieving.html (or from the Cambridge 
Crystallographic Data Centre, 12 Union Road, Cambridge CB2 1EZ, UK. Fax: +44-1223-336033. 
E-mail: deposit@ccdc.cam.ac.uk).

\subsection{Spectrum}
\label{sec:spect}

\begin{figure}
  \includegraphics[width=0.5\textwidth]{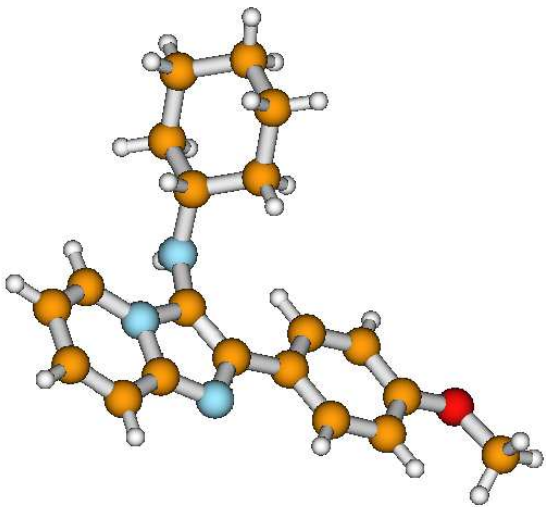}
  \caption{Experimental geometry: carbon, orange; nitrogen, blue; oxygen, red; hydrogen, white. This geometry consists of two nearly planar entities, namely a nearly planar cyclohexane (C$_6$H$_{11}$-) ring
and the rest of the molecule which rests in a plane perpendicular to the plane of the cyclohexane \label{fig:flugi6-exp-geo}}
\end{figure}
The experimental UV/Vis spectrum was determined in dimethyl sulphur dioxide (DMSO) as described
in the supplementary material of Ref. \cite{BMO11}
The first step to calculating the spectrum of Flugi {\bf 6} is to optimize the geometry.  This was done with {\sc BigDFT}
using the 0.4/8/8 grid,  
beginning with the experimental crystal geometry shown in Fig.~\ref{fig:flugi6-exp-geo}.
Our optimized
geometry is given in Table~\ref{tab:opt_Flugi_6}.
The largest change 0.15 {\AA} from the initial guess is for the 18th atom of hydrogen.
It was verified that the optimized geometry is indeed a minimum 
by explicit calculation of vibrational frequencies.  However the experimental geometry is not identical to our
calculated gas phase geometry as confirmed by the presence of three imaginary vibrational 
frequencies calculated for the (unoptimized) experimental geometry.

\begin{table} [!h]
  \caption{DFT optimized geometries (Cartesian coordinates in {\AA}) of Flugi {\bf 6}.
Calculations performed at the LDA level of theory. \label{tab:opt_Flugi_6}}
  \begin{tabular}{cccc}
  \hline \hline
  Atom &   $x$ & $y$  &  $z$ \\
  \hline
O        &      14.0374   &   1.5760 &  2.9779\\
N        &      7.75774   &   0.5820 &  7.4296\\
N        &      9.14347   &   2.5138 &  7.6339\\
N        &      8.73705   &   -0.581 &  5.7976\\
C        &      8.85269   &   1.3197 &  7.0273\\
C        &      7.73253   &   -0.574 &  6.6557\\
C        &      6.82855   &   0.8581 &  8.3749\\
C        &      5.83564   &   -0.044 &  8.6043\\
C        &      9.44155   &   0.5586 &  6.0148\\
C        &      12.4807   &   2.2541 &  4.5863\\
C        &      9.66483   &   2.5868 &  8.9771\\
C        &      12.2472   &   0.0755 &  3.6050\\
C        &      12.9348   &   1.2789 &  3.7031\\
C        &      11.3530   &   2.0325 &  5.3457\\
C        &      11.1111   &   -0.130 &  4.3645\\
C        &      10.6388   &   0.8366 &  5.2444\\
C        &      5.77383   &   -1.232 &  7.8551\\
C        &      6.70657   &   -1.492 &  6.8869\\
C        &      11.0671   &   2.0245 &  9.0831\\
C        &      14.5517   &   0.5791 &  2.1429\\
C        &      9.63546   &   4.0197 &  9.4485\\
C        &      10.1799   &   4.1615 &  10.852\\
C        &      11.5902   &   2.1260 &  10.497\\
C        &      11.5622   &   3.5584 &  10.985\\
H        &      6.94062   &   1.8172 &  8.8855\\
H        &      5.08962   &   0.1718 &  9.3696\\
H        &      13.0361   &   3.1924 &  4.6593\\
H        &      9.01856   &   1.9925 &  9.6670\\
H        &      12.5948   &   -0.707 &  2.9284\\
H        &      11.0089   &   2.7962 &  6.0486\\
H        &      10.5470   &   -1.064 &  4.2928\\
H        &      4.97202   &   -1.948 &  8.0454\\
H        &      6.68074   &   -2.397 &  6.2775\\
H        &      11.0746   &   0.9834 &  8.7183\\
H        &      11.7169   &   2.5999 &  8.3962\\
H        &      10.2520   &   4.6168 &  8.7465\\
H        &      8.60472   &   4.4110 &  9.3825\\
H        &      10.1918   &   5.2227 &  11.149\\
H        &      9.49548   &   3.6524 &  11.557\\
H        &      10.9656   &   1.5016 &  11.164\\
H        &      12.6112   &   1.7164 &  10.558\\
H        &      12.2802   &   4.1561 &  10.392\\
H        &      11.9013   &   3.6190 &  12.032\\
H        &      15.4016   &   1.0289 &  1.6155\\
H        &      14.9021   &   -0.292 &  2.7251\\
H        &      13.8030   &   0.2392 &  1.4055\\
H        &      8.94041   &   3.3910 &  7.1600\\
 \hline  \hline
  \end{tabular}
\end{table}

\begin{figure}
   \includegraphics[width=0.8\textwidth]{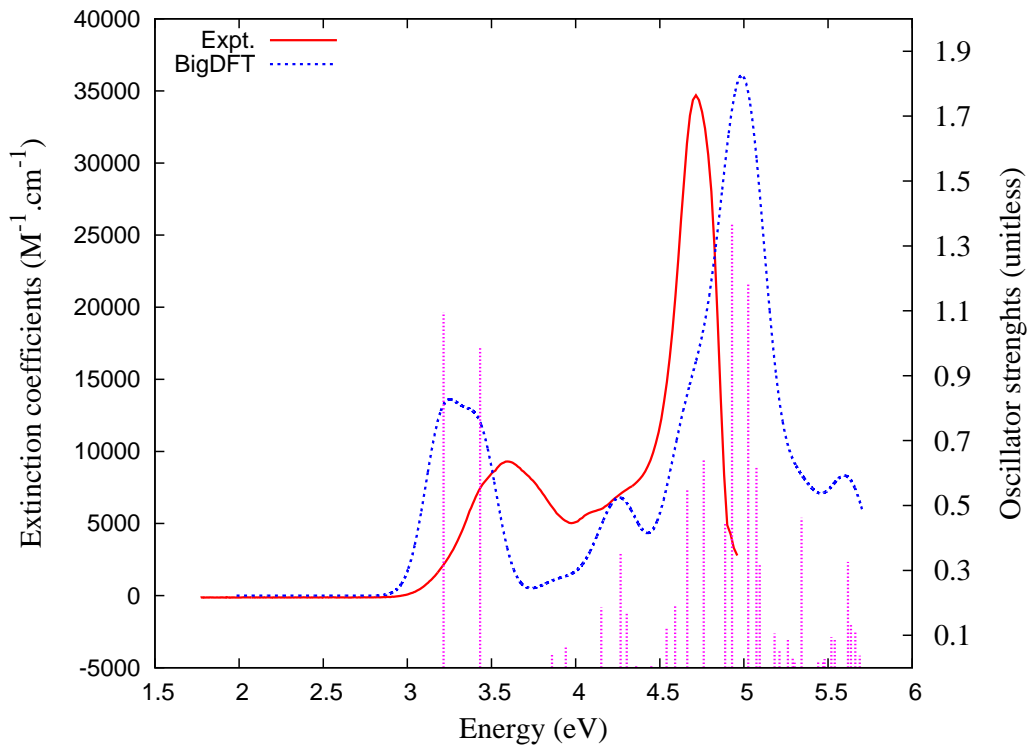}
   \caption{Comparison of theoretical and measured absorption spectra for Flugi {\bf 6} (left y-axis). The magnitude of the {\sc BigDFT} curve has been
divided by a factor of ten (see text). Both theoretical and experimental curves 
 show qualitative agreement with the oscillator strength stick-spectra which
however is in different units (right y-axis).
           \label{fig:flugi6-spectra}}
\end{figure}

The TD-LDA absorption spectrum of Flugi {\bf 6} was then calculated at the optimized geometry using our new implementation of 
TD-DFT in {\sc BigDFT}.  In addition to the previously mentioned computational details,
the calculation used 60 unoccupied orbitals within the TDA.
The excited-states were obtained using full-matrix diagonalization of the TD-DFT part.
The theoretical spectrum was calculated using Eq.~(\ref{eq:tddft.10d}) using a FWHM of 0.25 eV and then transformed to a
wavelength scale using our spectrum convolution program \cite{pablo}.  
Because we were restricted to a more limited number of unoccupied orbitals than in the N$_2$ test case, there is some concern that 
our calculated spectrum might change if a larger number of unoccupied orbitals is included.  However the comparison
of the theoretical and experimental results shown in Fig.~\ref{fig:flugi6-spectra} is reasonable.  This is 
especially true when it is kept in mind that we are comparing gas-phase theory with an experimental spectrum measured
in a polar solvent DMSO.  Notice the presence of a larger peak at 4.6-5.0 eV, a smaller peak at 3.2-3.5 eV,
and a shoulder inbetween near 4.3 eV.  


\begin{table} [!h]
  \caption{Singlet excitation energies ($\hbar \omega_I$, in eV) up to $-\epsilon_{\text{HOMO}}$ = 4.8713 eV, oscillator strength 
  ($f_I$, unitless) and assignment. \label{tab:flugi6-stick-spectrum}}
  \begin{tabular}{llll}
  \hline \hline
  $\hbar \omega_I$ & $f_I$ & Dominant transition\footnotemark[1] & Coefficient\footnotemark[2] \\
  \hline
  4.83006 & 0.047 & $^1(H,L+7)$ & 0.698419 \\ 
  4.81870 & 0.890 & $^1(H-2,L+3)$ & 0.402878 \\ 
  4.77478 & 0.054 & $^1(H,L+6)$ & 0.662945 \\ 
  4.72164 & 0.373 & $^1(H-1,L+4)$ & 0.423823 \\ 
  4.63424 & 0.665 & $^1(H-1,L+2)$ & 0.308510 \\ 
  4.58075 & 0.572 & $^1(H-1,L+3)$ & 0.279673 \\ 
  4.50552 & 0.048 & $^1(H,L+5)$ & 0.654877 \\ 
  4.47587 & 0.012 & $^1(H-5,L)$ & 0.396204 \\ 
  4.38722 & 0.037 & $^1(H-2,L+2)$ & 0.496849 \\ 
  4.33889 & 0.005 & $^1(H,L+4)$ & 0.550260 \\ 
  4.29785 & 0.007 & $^1(H-1,L)$ & 0.332720 \\ 
  4.24700 & 0.221 & $^1(H,L+3)$ & 0.418216 \\ 
  4.12777 & 0.495 & $^1(H,L+2)$ & 0.358160 \\ 
  3.92056 & 0.116 & $^1(H-2,L)$ & 0.431698 \\ 
  3.78973 & 0.042 &  $^1(H,L+2)$ & 0.418160 \\ 
  3.50331 & 0.883 &  $^1(H-1,L)$ & 0.523227 \\ 
  3.21284 & 1.386 &  $^1(H,L)$ & 0.615044 \\ 
  \hline \hline
  \end{tabular}
  \footnotetext[1]{$H$ and $L$ stand respectively for HOMO and LUMO.}
  \footnotetext[2]{Configuration interaction (CI) expansion coefficient.}
\end{table}

In contrast to experience with ruthenium complexes (to name but one 
example), Eq.(\ref{eq:tddft.10d}) with gaussian broadening does {\em not}
suffice in the present case to give good agreement between theoretical 
and experimental molar extinction coefficinets.  This is why the theoretical
curve has been divided {\em magnitude by a factor of ten}.  We believe that this
discrepancy is in part due to the aforementioned solvent effects on excitation
energies and oscillator strengths which we have chosen to neglect
and in part due to the possible presence of multiple conformers in
the room temperature experiment.  This latter hypothesis might be tested
by expensive dynamics calculations, but this far beyond the scope of 
the present work.  Nevertheless, even without dynamics we find this level of
agreement to be encouraging and now go on to further analyze our calculated spectrum.
The calculated stick spectrum is also shown in Fig.~\ref{fig:flugi6-spectra}.  It is now clear that the small peak
at 3.2-3.5 eV corresponds to two transitions, that the shoulder near 4.3 eV corresponds to three transitions,
and that the large peak at 4.6-5.0 eV corresponds to several electric excited states.  Table~\ref{tab:flugi6-stick-spectrum}
provides a more detailed analysis. All of the transitions are below the onset of TD-LDA ionization continuum
at $-\epsilon_{\text{HOMO}}$ = 4.8713 eV, which is artificially low compared to the true ionization
energy \cite{CJC+98}. 
As mentioned in Sec.~\ref{sec:BigDFT}, unlike TD 
Hartree-Fock (or configuration interaction singles) calculations, TD-LDA calculations are preprepared to 
describe excitation processes in the sense that the occupied and unoccupied orbitals both see the same number 
of electrons (because they come from the same local potential.)  This means that there is often little 
relaxation --- at least in small molecules --- and a two orbital model \cite{C95} provides a good first 
approximation to the excitation energy [Eqs.~(\ref{eq:BigDFT.14}) and (\ref{eq:BigDFT.15})].
The TDA configuration interaction coefficient is then determined by spin coupling and is given by
$1/\sqrt{2} = 0.707$.  Table~\ref{tab:flugi6-stick-spectrum} shows significant deviations from this
theoretical value, suggesting significant relaxation effects may be taking place.  Visualization
of the HOMO and LUMO suggests that relaxation is important here and might help to explain why the
first peak is at slightly too low an energy compared to the first experimental peak in the absorption
spectrum \cite{CGG+00}.  Nevertheless, the energy of HOMO $\rightarrow$ LUMO dominated
singlet transition at 3.21 eV exceeds the simple difference of HOMO and LUMO molecular orbital energies
of 2.80 eV as expected from the domination of the Hartree term $(ia \vert f_H \vert ai)$ over the
two xc terms. 
Two interesting features, which will not
be pursued in the present paper, are the absence of oscillator strength for the $^1(H,L+1)$ transition 
and the indication of significant configuration mixing for the first and seventh transitions which 
both borrow from $^1(H,L)$ and for the third and fifth transitions which both borrow from $^1(H,L+2)$.



\section{Conclusion}
\label{sec:conclude}

Grid-based methods have long been regarded with skepticism by the quantum chemistry community, but 
have now been accepted in the form of the grids used to evaluate xc-integrals in the DFT part of
most quantum chemistry codes.  We believe that an even greater acceptance of grid-based methods may
prove useful as theoretical solid-state and chemical physics strive to meet on ``neutral ground''
at the nanometer scale.  Acceptance will be aided by continuing advances in grid-based methods 
with wavelets being of particular interest here.  At the same time, new features should be added 
to grid-based electronic structure codes in order to make them more useful for chemical applications.  
This paper represents a small step in that direction.  In particular, we have presented the first 
implementation of TD-DFT in a wavelet-based code, namely in {\sc BigDFT}.

While {\sc BigDFT} is designed for routine calculations on systems containing many hundreds of
atoms, this first implementation of TD-DFT in {\sc BigDFT} is not yet ready for these more
ambitious applications.  Rather, we wished to bring out the pros and cons of wavelet-based
TD-DFT by comparing against a GTO-based quantum chemistry code (in this case, {\sc deMon2k}) and by an
example application showing how our implementation in {\sc BigDFT} can be useful in analyzing
the spectrum of a molecule of contemporary experimental interest.

A factor in favor of the wavelet-based approach is the rapidity of convergence of the
bound orbitals and orbital energies with respect to refinements of the wavelet basis
and associated grid.  While orbital results from the all-electron {\sc deMon2k} code 
are not (and should not) be in exact agreement with those of the pseudopotential {\sc BigDFT} 
code, the results are really quite close when sufficiently large basis sets are used.
In the case of the GTO-based code, tight basis set convergence typically requires
going beyond at least the triple-zeta-valence-plus-polarization (TZVP) level.  In contrast, 
adequate orbital convergence is easily obtained with {\sc BigDFT} using the default wavelet basis
set and grid, with further refinements leading to only minor improvement.  This comes
close to the quantum chemists' dream of calculations free of errors due to basis set
incompleteness.

Interestingly, problems which could be envisioned with this first implementation of TD-DFT in {\sc BigDFT}
either did not arise or did not seem to be serious.  The worry was that the unbound orbitals of the 
molecule are continuum orbitals whose description is apparently
only limited by the boundaries of the box defined by the coarse grid.  In principle, for
an infinitely large box, there are an infinite number of unoccupied orbitals in even
a small energy band and all of these would seem to need to be taken into account even for describing transitions below the TD-DFT ionization threshold at -$\epsilon_{HOMO}$.  This
is a doubly large worry because the number of unoccupied orbitals to be calculated 
is  limited by an input parameter, making a double convergence problem (number of virtuals and
box size.)  Nevertheless our calculations show that the 
implementation in {\sc BigDFT} works correctly, giving quite reasonable results 
when compared with {\sc deMon2k} and with experimental results for transitions
below -$\epsilon_{HOMO}$.  There are at least
two probable reasons that the anticipated problems are not seen here.  The first is the
tendency, at least for small molecules, for excitations to be dominated by bound-to-bound
transitions involving two or only a few orbitals.  This is especially true for the LDA
and GGA, but will gradually breakdown with the inclusion of Hartree-Fock exchange where
orbital relaxation becomes more important.
The second reason that we may not see the expected problems associated with continuum-type
unoccupied orbitals is that putting the molecule in a box acts much like atom-centered 
GTOs in the sense that it keeps wavelet basis functions close to those regions of space 
where electron density is high and so can be efficiently used to describe the dynamic 
response of the charge density, whose description is the fundamental key to extracting 
spectra in TD-DFT.  Nevertheless
it should be mentioned that there are alternative algorithms in TD-DFT such as the
modified Sternheimer equation and the Green's function approach which avoid explicit 
reference to unoccupied orbitals \cite{MS90}.  These may be worth exploring in future
development work of wavelet-based TD-DFT.  However our first priority will be to implement
analytic derivatives for TD-DFT excited states which are needed in modeling fluorescence.

It should also be mentioned that implementing TD-DFT is a step along the way to
implementing MBPT methods from solid-state physics, namely the $GW$ one-particle
Green's function approximation and the Bethe-Salpeter equation approach to the
two-particle Green's function \cite{ORR02}.  Such work is already in progress in
{\sc BigDFT}. Of course, the expected collapse of the TD-DFT continuum \cite{CJC+98} was seen above -$\epsilon_{HOMO}$ (Fig. \ref{fig:excited-states-continuum}) with increasing box size, though the spectrum remains qualitatively
correct (Fig. \ref{fig:n2-spectra-bigdft-demon-0}).

The application to Flugi {\bf 6} presented in this paper provides a concrete reality check
on the usefulness of TD-DFT for practical applications.  Geometry optimization was
simplified by beginning with an x-ray structure, but solvent and dynamics effects were ignored.
All in all the final result may be qualified as semiquantitative but useful.
In fact, we have also carried out preliminary calculations of absorption spectra for five 
other members of the Flugi combinatorial chemistry series (not reported here.)  For 
these molecules, trends in the energies of the first experimental absorption peaks do 
parallel trends in the calculated first absorption peak as well as the HOMO-LUMO energy 
difference.  However, in the absence of x-ray crystal data, the large number of possible 
molecular configurations merits further exploration, especially since the LDA may not 
correctly order these states.  Hydrogen bonding with the solvent should also be considered
in some cases.  For these reasons it seems best to reserve the calculation of these spectra, 
comparison with experimental spectra, and assignment of transitions for a future paper.

 
\begin{acknowledgments}

B.\ N.\ would like to acknowledge a scholarship from the {\em Fondation Nanosciences}.
Those of us at the {\em Universit\'e Joseph Fourier} would like to thank
Denis Charapoff, R\'egis Gras, S\'ebastien Morin, and Marie-Louise Dheu-Andries for technical
support at the (DCM) and for technical support in the context of the {\em Centre d'Exp\'erimentation
du Calcul Intensif en Chimie} (CECIC) computers used for some of the calculations reported here.
This work has been carried out in the context of the French Rh\^one-Alpes
{\em R\'eseau th\'ematique de recherche avanc\'ee (RTRA): Nanosciences aux limites de la 
nano\'electronique} and the Rh\^one-Alpes Associated Node of the European Theoretical 
Spectroscopy Facility (ETSF).


\end{acknowledgments}
\newpage
 \singlespace
\bibliographystyle{myaip}
\bibliography{refs}

\begin{thebibliography}{10}%
\makeatletter
\providecommand \@ifxundefined [1]{%
 \ifx #1\undefined \expandafter \@firstoftwo
 \else \expandafter \@secondoftwo
\fi
}%
\providecommand \@ifnum [1]{%
 \ifnum #1\expandafter \@firstoftwo
 \else \expandafter \@secondoftwo
\fi
}%
\providecommand \enquote [1]{``#1''}%
\providecommand \bibnamefont  [1]{#1}%
\providecommand \bibfnamefont [1]{#1}%
\providecommand \citenamefont [1]{#1}%
\providecommand\href[0]{\@sanitize\@href}%
\providecommand\@href[1]{\endgroup\@@startlink{#1}\endgroup\@@href}%
\providecommand\@@href[1]{#1\@@endlink}%
\providecommand \@sanitize [0]{\begingroup\catcode`\&12\catcode`\#12\relax}%
\@ifxundefined \pdfoutput {\@firstoftwo}{%
 \@ifnum{\z@=\pdfoutput}{\@firstoftwo}{\@secondoftwo}%
}{%
 \providecommand\@@startlink[1]{\leavevmode\special{html:<a href="#1">}}%
 \providecommand\@@endlink[0]{\special{html:</a>}}%
}{%
 \providecommand\@@startlink[1]{%
  \leavevmode
  \pdfstartlink
   attr{/Border[0 0 1 ]/H/I/C[0 1 1]}%
   user{/Subtype/Link/A<</Type/Action/S/URI/URI(#1)>>}%
  \relax
 }%
 \providecommand\@@endlink[0]{\pdfendlink}%
}%
\providecommand \url  [0]{\begingroup\@sanitize \@url }%
\providecommand \@url [1]{\endgroup\@href {#1}{\urlprefix}}%
\providecommand \urlprefix [0]{URL }%
\providecommand \Eprint[0]{\href }%
\@ifxundefined \urlstyle {%
  \providecommand \doi [1]{doi:\discretionary{}{}{}#1}%
}{%
  \providecommand \doi [0]{doi:\discretionary{}{}{}\begingroup
  \urlstyle{rm}\Url }%
}%
\providecommand \doibase [0]{http://dx.doi.org/}%
\providecommand \Doi[1]{\href{\doibase#1}}%
\providecommand \bibAnnote [3]{%
  \BibitemShut{#1}%
  \begin{quotation}\noindent
    \textsc{Key:}\ #2\\\textsc{Annotation:}\ #3%
  \end{quotation}%
}%
\providecommand \bibAnnoteFile [2]{%
  \IfFileExists{#2}{\bibAnnote {#1} {#2} {\input{#2}}}{}%
}%
\providecommand \typeout [0]{\immediate \write \m@ne }%
\providecommand \selectlanguage [0]{\@gobble}%
\providecommand \bibinfo [0]{\@secondoftwo}%
\providecommand \bibfield [0]{\@secondoftwo}%
\providecommand \translation [1]{[#1]}%
\providecommand \BibitemOpen[0]{}%
\providecommand \bibitemStop [0]{}%
\providecommand \bibitemNoStop [0]{.\EOS\space}%
\providecommand \EOS [0]{\spacefactor3000\relax}%
\providecommand \BibitemShut [1]{\csname bibitem#1\endcsname}%
\bibitem{A99}%
  \BibitemOpen
  \bibfield{author}{%
  \bibinfo {author} {\bibfnamefont{T.~A.}\ \bibnamefont{Arias}},\ }%
  \bibfield{journal}{%
  \bibinfo {journal} {Rev. Mod. Phys.}\ }%
  \textbf{\bibinfo {volume} {71}},\ \bibinfo {pages} {267} (\bibinfo {year}
  {1999}),\ \bibinfo {note} {{\sf Multiresolution analysis of electronic
  structure: semicardinal and wavelet bases}}%
  \bibAnnoteFile{NoStop}{A99}%
\bibitem{G98}%
  \BibitemOpen
  \bibfield{author}{%
  \bibinfo {author} {\bibfnamefont{S.}~\bibnamefont{Goedecker}},\ }%
  \emph{\bibinfo {title} {Wavelets and Their Application for the Solution of
  Partial Differential Equations}}\ (\bibinfo {publisher} {Presses
  Polytechniques Universitaires et Romandes},\ \bibinfo {address} {Lausanne,
  Switzerland},\ \bibinfo {year} {1998})%
  \bibAnnoteFile{NoStop}{G98}%
\bibitem{bigdft}%
  \BibitemOpen
  \bibinfo {howpublished} {\url{http://inac.cea.fr/L_Sim/BigDFT/}}%
  \bibAnnoteFile{NoStop}{bigdft}%
\bibitem{deMon2k}%
  \BibitemOpen
  \bibinfo {note} {{\sc deMon2k{@}Grenoble}, the Grenoble development version
  of {\sc deMon2k}, Andreas M.\ K\"oster, Patrizia Calaminici, Mark E.\ Casida,
  Roberto Flores-Morino, Gerald Geudtner, Annick Goursot, Thomas Heine, Andrei
  Ipatov, Florian Janetzko, Sergei Patchkovskii, J.\ Ulisis Reveles, Dennis R.
  Salahub, and Alberto Vela, {\em The International deMon Developers Community}
  (Cinvestav-IPN, Mexico, 2006) {\em plus some additional features} by Mark E.
  Casida, Lo\"{\i}c Joubert Doriol, Andrei Ipatov, Miquel Huix-Rotllant, and
  Bhaarathi Natarajan (Grenoble, France, 2011).}%
  \bibAnnoteFile{Stop}{deMon2k}%
\bibitem{BMO11}%
  \BibitemOpen
  \bibfield{author}{%
  \bibinfo {author} {\bibfnamefont{O.~N.}\ \bibnamefont{Burchak}}, \bibinfo
  {author} {\bibfnamefont{L.}~\bibnamefont{Mugherli}}, \bibinfo {author}
  {\bibfnamefont{M.}~\bibnamefont{Ostuni}}, \bibinfo {author}
  {\bibfnamefont{J.~J.}\ \bibnamefont{Lacapère}},\ and\ \bibinfo {author}
  {\bibfnamefont{M.~Y.}\ \bibnamefont{Balakirev}},\ }%
  \bibfield{journal}{%
  \bibinfo {journal} {Journal of the American Chemical Society}\ }%
  \textbf{\bibinfo {volume} {133}},\ \bibinfo {pages} {10058} (\bibinfo {year}
  {2011}),\ \bibinfo {note} {{\sf Combinatorial Discovery of Fluorescent
  Pharmacophores by Multicomponent Reactions in Droplet Arrays}}%
  \bibAnnoteFile{NoStop}{BMO11}%
\bibitem{abinitio}%
  \BibitemOpen
  \bibinfo {note} {Note that DFT is usually considered as {\em ab initio} in
  solid-state theory, while chemical physicists often make a distinction
  between {\em ab initio} and DFT.}%
  \bibAnnoteFile{Stop}{abinitio}%
\bibitem{CKSL08}%
  \BibitemOpen
  \bibfield{author}{%
  \bibinfo {author} {\bibfnamefont{C.}~\bibnamefont{Bienia}}, \bibinfo {author}
  {\bibfnamefont{S.}~\bibnamefont{Kumar}}, \bibinfo {author}
  {\bibfnamefont{J.~P.}\ \bibnamefont{Singh}},\ and\ \bibinfo {author}
  {\bibfnamefont{K.}~\bibnamefont{Li}},\ }%
  \emph{\bibinfo {title} {The {PARSEC} Benchmark Suite: Characterization and
  Architectural Implications}},\ \bibinfo {type} {Tech. Rep.}\ (\bibinfo
  {institution} {in Princeton University},\ \bibinfo {year} {2008})%
  \bibAnnoteFile{NoStop}{CKSL08}%
\bibitem{CMA+06}%
  \BibitemOpen
  \bibfield{author}{%
  \bibinfo {author} {\bibfnamefont{A.}~\bibnamefont{Castro}}, \bibinfo {author}
  {\bibfnamefont{M.~A.~L.}\ \bibnamefont{Marques}}, \bibinfo {author}
  {\bibfnamefont{H.}~\bibnamefont{Appel}}, \bibinfo {author}
  {\bibfnamefont{M.}~\bibnamefont{Oliveira}}, \bibinfo {author}
  {\bibfnamefont{C.~A.}\ \bibnamefont{Rozzi}}, \bibinfo {author}
  {\bibfnamefont{X.}~\bibnamefont{Andrade}}, \bibinfo {author}
  {\bibfnamefont{F.}~\bibnamefont{Lorenzen}}, \bibinfo {author}
  {\bibfnamefont{E.~K.~U.}\ \bibnamefont{Gross}},\ and\ \bibinfo {author}
  {\bibfnamefont{A.}~\bibnamefont{Rubio}},\ }%
  \bibfield{journal}{%
  \bibinfo {journal} {Physica Status Solidi}\ }%
  \textbf{\bibinfo {volume} {243}},\ \bibinfo {pages} {2465} (\bibinfo {year}
  {2006}),\ \bibinfo {note} {{\sf {\sc Octopus}: a tool for the application of
  time-dependent density functional theory}}%
  \bibAnnoteFile{NoStop}{CMA+06}%
\bibitem{MJ94}%
  \BibitemOpen
  \emph{\bibinfo {title} {Wavelet Applications in Chemical Engineering}},\
  edited by\ \bibinfo {editor} {\bibfnamefont{R.~L.}\ \bibnamefont{Motard}}\
  and\ \bibinfo {editor} {\bibfnamefont{B.}~\bibnamefont{Joseph}}\ (\bibinfo
  {publisher} {Kluwer Academicc Publishers},\ \bibinfo {address} {Boston},\
  \bibinfo {year} {1994})%
  \bibAnnoteFile{NoStop}{MJ94}%
\bibitem{C61}%
  \BibitemOpen
  \bibfield{author}{%
  \bibinfo {author} {\bibfnamefont{J.~W.}\ \bibnamefont{Cooley}},\ }%
  \bibfield{journal}{%
  \bibinfo {journal} {Math. Computation}\ }%
  \textbf{\bibinfo {volume} {15}},\ \bibinfo {pages} {363} (\bibinfo {year}
  {1961}),\ \bibinfo {note} {{\sf An Improved Eigenvalue Corrector Formula for
  Solving the Schr\"{o}dinger Equation for Central Fields}}%
  \bibAnnoteFile{NoStop}{C61}%
\bibitem{C63}%
  \BibitemOpen
  \bibfield{author}{%
  \bibinfo {author} {\bibfnamefont{J.~K.}\ \bibnamefont{Cashion}},\ }%
  \bibfield{journal}{%
  \bibinfo {journal} {J. Chem. Phys.}\ }%
  \textbf{\bibinfo {volume} {39}},\ \bibinfo {pages} {1872} (\bibinfo {year}
  {1963}),\ \bibinfo {note} {{\sf Testing of Diatomic Potential-Energy
  Functions by Numerical Methods}}%
  \bibAnnoteFile{NoStop}{C63}%
\bibitem{Z64}%
  \BibitemOpen
  \bibfield{author}{%
  \bibinfo {author} {\bibfnamefont{R.~N.}\ \bibnamefont{Zare}},\ }%
  \bibfield{journal}{%
  \bibinfo {journal} {J. Chem. Phys.}\ }%
  \textbf{\bibinfo {volume} {40}},\ \bibinfo {pages} {1934} (\bibinfo {year}
  {1964}),\ \bibinfo {note} {{\sf Calculation of Intensity Distribution in the
  Vibrational Structure of Electronic Transitions: {T}he {$B ^3\Pi_{O^+ u} - X
  ^1\Sigma_{O^+ g}$} Resonance Series of Molecular Iodine}}%
  \bibAnnoteFile{NoStop}{Z64}%
\bibitem{F72}%
  \BibitemOpen
  \bibfield{author}{%
  \bibinfo {author} {\bibfnamefont{C.~F.}\ \bibnamefont{Fischer}},\ }%
  \bibfield{journal}{%
  \bibinfo {journal} {At. Data Nucl. Data Tables}\ }%
  \textbf{\bibinfo {volume} {4}},\ \bibinfo {pages} {301} (\bibinfo {year}
  {1972}),\ \bibinfo {note} {{\sf Average-energy-of-configuration Hartree-Fock
  results for the atoms helium to radon charlotte froese fischer}}%
  \bibAnnoteFile{NoStop}{F72}%
\bibitem{M73}%
  \BibitemOpen
  \bibfield{author}{%
  \bibinfo {author} {\bibfnamefont{J.~B.}\ \bibnamefont{Mann}}\ and\ \bibinfo
  {author} {\bibfnamefont{J.~T.}\ \bibnamefont{Waber}},\ }%
  \bibfield{journal}{%
  \bibinfo {journal} {At. Data Nucl. Data Tables}\ }%
  \textbf{\bibinfo {volume} {5}},\ \bibinfo {pages} {201} (\bibinfo {year}
  {1973}),\ \bibinfo {note} {{\sf Self-consistent relativistic
  Dirac-Hartree-Fock calculations of lanthanide atoms}}%
  \bibAnnoteFile{NoStop}{M73}%
\bibitem{F73}%
  \BibitemOpen
  \bibfield{author}{%
  \bibinfo {author} {\bibfnamefont{C.~F.}\ \bibnamefont{Fischer}},\ }%
  \bibfield{journal}{%
  \bibinfo {journal} {At. Data Nucl. Data Tables}\ }%
  \textbf{\bibinfo {volume} {12}},\ \bibinfo {pages} {87} (\bibinfo {year}
  {1973}),\ \bibinfo {note} {{\sf Average-energy-of-configuration Hartree-Fock
  results for the atoms helium to radon}}%
  \bibAnnoteFile{NoStop}{F73}%
\bibitem{D92}%
  \BibitemOpen
  \bibfield{author}{%
  \bibinfo {author} {\bibfnamefont{I.}~\bibnamefont{Daubechies}},\ }%
  \emph{\bibinfo {title} {Ten Lectures on Wavelets}},\ \bibinfo {series}
  {CBMS-NSF}, Vol.~\bibinfo {volume} {61}\ (\bibinfo {publisher} {SIAM},\
  \bibinfo {address} {Philadelphia},\ \bibinfo {year} {1992})%
  \bibAnnoteFile{NoStop}{D92}%
\bibitem{D96}%
  \BibitemOpen
  \bibfield{author}{%
  \bibinfo {author} {\bibfnamefont{I.}~\bibnamefont{Daubechies}},\ }%
  \bibfield{journal}{%
  \bibinfo {journal} {Proceedings of the IEEE}\ }%
  \textbf{\bibinfo {volume} {84}},\ \bibinfo {pages} {510} (\bibinfo {year}
  {1996}),\ \bibinfo {note} {{\sf Where do wavelets come from? --- {A} personal
  point of view}}%
  \bibAnnoteFile{NoStop}{D96}%
\bibitem{GM10}%
  \BibitemOpen
  \bibfield{author}{%
  \bibinfo {author} {\bibfnamefont{S.}~\bibnamefont{Gopalakrishnan}}\ and\
  \bibinfo {author} {\bibfnamefont{M.}~\bibnamefont{Mitra}},\ }%
  \emph{\bibinfo {title} {Wavelet Methods for Dynamical Problems}}\ (\bibinfo
  {publisher} {CRC Press},\ \bibinfo {address} {New York},\ \bibinfo {year}
  {2010})%
  \bibAnnoteFile{NoStop}{GM10}%
\bibitem{FD93}%
  \BibitemOpen
  \bibfield{author}{%
  \bibinfo {author} {\bibfnamefont{P.}~\bibnamefont{Fischer}}\ and\ \bibinfo
  {author} {\bibfnamefont{M.}~\bibnamefont{Defranceschi}},\ }%
  \bibfield{journal}{%
  \bibinfo {journal} {Int. J. Quant. chem.}\ }%
  \textbf{\bibinfo {volume} {45}},\ \bibinfo {pages} {619} (\bibinfo {year}
  {1993}),\ \bibinfo {note} {{\sf Looking at atomic orbitals through {F}ourier
  and wavelet transforms}}%
  \bibAnnoteFile{NoStop}{FD93}%
\bibitem{C96b}%
  \BibitemOpen
  \bibfield{author}{%
  \bibinfo {author} {\bibfnamefont{J.-L.}\ \bibnamefont{Calais}},\ }%
  \bibfield{journal}{%
  \bibinfo {journal} {Int. J. Quant. Chem.}\ }%
  \textbf{\bibinfo {volume} {58}},\ \bibinfo {pages} {541} (\bibinfo {year}
  {1996}),\ \bibinfo {note} {{\sf Wavelets--{S}omething for quantum
  chemistry?}}%
  \bibAnnoteFile{Stop}{C96b}%
\bibitem{HK64}%
  \BibitemOpen
  \bibfield{author}{%
  \bibinfo {author} {\bibfnamefont{P.}~\bibnamefont{Hohenberg}}\ and\ \bibinfo
  {author} {\bibfnamefont{W.}~\bibnamefont{Kohn}},\ }%
  \bibfield{journal}{%
  \bibinfo {journal} {Phys. Rev.}\ }%
  \textbf{\bibinfo {volume} {136}},\ \bibinfo {pages} {B864} (\bibinfo {year}
  {1964}),\ \bibinfo {note} {{\sf {I}nhomogeneous electron gas}}%
  \bibAnnoteFile{NoStop}{HK64}%
\bibitem{KS65}%
  \BibitemOpen
  \bibfield{author}{%
  \bibinfo {author} {\bibfnamefont{W.}~\bibnamefont{Kohn}}\ and\ \bibinfo
  {author} {\bibfnamefont{L.~J.}\ \bibnamefont{Sham}},\ }%
  \bibfield{journal}{%
  \bibinfo {journal} {Phys. Rev. A}\ }%
  \textbf{\bibinfo {volume} {140}},\ \bibinfo {pages} {1133} (\bibinfo {year}
  {1965}),\ \bibinfo {note} {{\sf {S}elf-consistent equations including
  exchange-correlation effects}}%
  \bibAnnoteFile{NoStop}{KS65}%
\bibitem{HFYB03}%
  \BibitemOpen
  \bibfield{author}{%
  \bibinfo {author} {\bibfnamefont{R.~J.}\ \bibnamefont{Harrison}}, \bibinfo
  {author} {\bibfnamefont{G.~I.}\ \bibnamefont{Fann}}, \bibinfo {author}
  {\bibfnamefont{T.}~\bibnamefont{Yanai}},\ and\ \bibinfo {author}
  {\bibfnamefont{G.}~\bibnamefont{Beylkin}},\ }%
  in\ \emph{\bibinfo {booktitle} {ICCS 2003, LNCS 2660}},\ \bibinfo {editor}
  {edited by\ \bibinfo {editor} {\bibfnamefont{P.~M. A.~S.}\ \bibnamefont{{\em
  et al.}}}}\ (\bibinfo {publisher} {Springer-Verlag},\ \bibinfo {address}
  {Berlin},\ \bibinfo {year} {2003})\ pp.\ \bibinfo {pages} {103--110},\
  \bibinfo {note} {{\sf Multiresolution quantum chemistry in multiwavelet
  bases}}%
  \bibAnnoteFile{NoStop}{HFYB03}%
\bibitem{RG84}%
  \BibitemOpen
  \bibfield{author}{%
  \bibinfo {author} {\bibfnamefont{E.}~\bibnamefont{Runge}}\ and\ \bibinfo
  {author} {\bibfnamefont{E.~K.~U.}\ \bibnamefont{Gross}},\ }%
  \bibfield{journal}{%
  \bibinfo {journal} {Phys. Rev. Lett.}\ }%
  \textbf{\bibinfo {volume} {52}},\ \bibinfo {pages} {997} (\bibinfo {year}
  {1984}),\ \bibinfo {note} {{\sf {D}ensity-functional theory for
  time-dependent systems}}%
  \bibAnnoteFile{NoStop}{RG84}%
\bibitem{C95}%
  \BibitemOpen
  \bibfield{author}{%
  \bibinfo {author} {\bibfnamefont{M.~E.}\ \bibnamefont{Casida}},\ }%
  in\ \emph{\bibinfo {booktitle} {Recent Advances in Density Functional
  Methods, Part I}},\ \bibinfo {editor} {edited by\ \bibinfo {editor}
  {\bibfnamefont{D.~P.}\ \bibnamefont{Chong}}}\ (\bibinfo {publisher} {World
  Scientific},\ \bibinfo {address} {Singapore},\ \bibinfo {year} {1995})\ p.\
  \bibinfo {pages} {155},\ \bibinfo {note} {{\sf {T}ime-dependent
  density-functional response theory for molecules}}%
  \bibAnnoteFile{NoStop}{C95}%
\bibitem{JCS96}%
  \BibitemOpen
  \bibfield{author}{%
  \bibinfo {author} {\bibfnamefont{C.}~\bibnamefont{Jamorski}}, \bibinfo
  {author} {\bibfnamefont{M.~E.}\ \bibnamefont{Casida}},\ and\ \bibinfo
  {author} {\bibfnamefont{D.~R.}\ \bibnamefont{Salahub}},\ }%
  \bibfield{journal}{%
  \bibinfo {journal} {J. Chem. Phys.}\ }%
  \textbf{\bibinfo {volume} {104}},\ \bibinfo {pages} {5134} (\bibinfo {year}
  {1996}),\ \bibinfo {note} {{\sf {D}ynamic polarizabilities and excitation
  spectra from a molecular implementation of time-dependent density-functional
  response theory: {N}$_2$ as a case study}}%
  \bibAnnoteFile{NoStop}{JCS96}%
\bibitem{CJC+98}%
  \BibitemOpen
  \bibfield{author}{%
  \bibinfo {author} {\bibfnamefont{M.~E.}\ \bibnamefont{Casida}}, \bibinfo
  {author} {\bibfnamefont{C.}~\bibnamefont{Jamorski}}, \bibinfo {author}
  {\bibfnamefont{K.~C.}\ \bibnamefont{Casida}},\ and\ \bibinfo {author}
  {\bibfnamefont{D.~R.}\ \bibnamefont{Salahub}},\ }%
  \bibfield{journal}{%
  \bibinfo {journal} {J. Chem. Phys.}\ }%
  \textbf{\bibinfo {volume} {108}},\ \bibinfo {pages} {4439} (\bibinfo {year}
  {1998}),\ \bibinfo {note} {{\sf {M}olecular excitation energies to high-lying
  bound states from time-dependent density-functional response theory:
  {C}haracterization and correction of the time-dependent local density
  approximation ionization threshold}}%
  \bibAnnoteFile{NoStop}{CJC+98}%
\bibitem{GK90}%
  \BibitemOpen
  \bibfield{author}{%
  \bibinfo {author} {\bibfnamefont{E.~K.~U.}\ \bibnamefont{Gross}}\ and\
  \bibinfo {author} {\bibfnamefont{W.}~\bibnamefont{Kohn}},\ }%
  \bibfield{journal}{%
  \bibinfo {journal} {Adv. Quant. Chem.}\ }%
  \textbf{\bibinfo {volume} {21}},\ \bibinfo {pages} {255} (\bibinfo {year}
  {1990}),\ \bibinfo {note} {{\sf {T}ime-dependent density functional theory}}%
  \bibAnnoteFile{NoStop}{GK90}%
\bibitem{GUG94}%
  \BibitemOpen
  \bibfield{author}{%
  \bibinfo {author} {\bibfnamefont{E.~K.~U.}\ \bibnamefont{Gross}}, \bibinfo
  {author} {\bibfnamefont{C.~A.}\ \bibnamefont{Ullrich}},\ and\ \bibinfo
  {author} {\bibfnamefont{U.~J.}\ \bibnamefont{Gossmann}},\ }%
  in\ \emph{\bibinfo {booktitle} {Density Functional Theory}},\ \bibinfo
  {editor} {edited by\ \bibinfo {editor} {\bibfnamefont{E.~K.~U.}\
  \bibnamefont{Gross}}\ and\ \bibinfo {editor} {\bibfnamefont{R.~M.}\
  \bibnamefont{Dreizler}}}\ (\bibinfo {publisher} {Plenum},\ \bibinfo {address}
  {New York},\ \bibinfo {year} {1994})\ pp.\ \bibinfo {pages} {149--171},\
  \bibinfo {note} {{\sf {D}ensity functional theory of time-dependent
  systems}}%
  \bibAnnoteFile{NoStop}{GUG94}%
\bibitem{GDP96}%
  \BibitemOpen
  \bibfield{author}{%
  \bibinfo {author} {\bibfnamefont{E.~K.~U.}\ \bibnamefont{Gross}}, \bibinfo
  {author} {\bibfnamefont{J.~F.}\ \bibnamefont{Dobson}},\ and\ \bibinfo
  {author} {\bibfnamefont{M.}~\bibnamefont{Petersilka}},\ }%
  \bibfield{journal}{%
  \bibinfo {journal} {Topics in Current Chemistry}\ }%
  \textbf{\bibinfo {volume} {181}},\ \bibinfo {pages} {81} (\bibinfo {year}
  {1996}),\ \bibinfo {note} {{\sf {D}ensity-functional theory of time-dependent
  phenomena}}%
  \bibAnnoteFile{NoStop}{GDP96}%
\bibitem{C96}%
  \BibitemOpen
  \bibfield{author}{%
  \bibinfo {author} {\bibfnamefont{M.~E.}\ \bibnamefont{Casida}},\ }%
  in\ \emph{\bibinfo {booktitle} {Recent Developments and Applications of
  Modern Density Functional Theory}},\ \bibinfo {editor} {edited by\ \bibinfo
  {editor} {\bibfnamefont{J.~M.}\ \bibnamefont{Seminario}}}\ (\bibinfo
  {publisher} {Elsevier},\ \bibinfo {address} {Elsevier, Amsterdam},\ \bibinfo
  {year} {1996})\ p.\ \bibinfo {pages} {391},\ \bibinfo {note} {{\sf
  {T}ime-Dependent Density Functional Response Theory of Molecular Systems:
  {T}heory, Computational Methods, and Functionals}}%
  \bibAnnoteFile{NoStop}{C96}%
\bibitem{BG98}%
  \BibitemOpen
  \bibfield{author}{%
  \bibinfo {author} {\bibfnamefont{K.}~\bibnamefont{Burke}}\ and\ \bibinfo
  {author} {\bibfnamefont{E.~K.~U.}\ \bibnamefont{Gross}},\ }%
  in\ \emph{\bibinfo {booktitle} {Density Functionals: Theory and
  Applications}},\ \bibinfo {series} {Springer Lecture Notes in Physics}, Vol.\
  \bibinfo {volume} {500},\ \bibinfo {editor} {edited by\ \bibinfo {editor}
  {\bibfnamefont{D.}~\bibnamefont{Joubert}}}\ (\bibinfo {publisher}
  {Springer},\ \bibinfo {year} {1998})\ pp.\ \bibinfo {pages} {116--146},\
  \bibinfo {note} {{\sf {A} guided tour of time-dependent density functional
  theory}}%
  \bibAnnoteFile{NoStop}{BG98}%
\bibitem{L01}%
  \BibitemOpen
  \bibfield{author}{%
  \bibinfo {author} {\bibfnamefont{R.}~\bibnamefont{van Leeuwen}},\ }%
  \bibfield{journal}{%
  \bibinfo {journal} {Int. J. Mod. Phys. B}\ }%
  \textbf{\bibinfo {volume} {15}},\ \bibinfo {pages} {1969} (\bibinfo {year}
  {2001}),\ \bibinfo {note} {{\sf {K}ey concepts in time-dependent
  density-functional theory}}%
  \bibAnnoteFile{NoStop}{L01}%
\bibitem{VBB+01}%
  \BibitemOpen
  \bibfield{author}{%
  \bibinfo {author} {\bibfnamefont{G.}~\bibnamefont{te~Velde}}, \bibinfo
  {author} {\bibfnamefont{F.~M.}\ \bibnamefont{Bickelhaupt}}, \bibinfo {author}
  {\bibfnamefont{E.~J.}\ \bibnamefont{Baerends}}, \bibinfo {author}
  {\bibfnamefont{C.}~\bibnamefont{Fonseca}}, \bibinfo {author}
  {\bibfnamefont{C.}~\bibnamefont{Guerra}}, \bibinfo {author}
  {\bibfnamefont{S.~J.~A.}\ \bibnamefont{van Gisbergen}}, \bibinfo {author}
  {\bibfnamefont{J.~G.}\ \bibnamefont{Snijders}},\ and\ \bibinfo {author}
  {\bibfnamefont{T.}~\bibnamefont{Ziegler}},\ }%
  \bibfield{journal}{%
  \bibinfo {journal} {J. Comput. Chem.}\ }%
  \textbf{\bibinfo {volume} {22}},\ \bibinfo {pages} {931} (\bibinfo {year}
  {2001}),\ \bibinfo {note} {{\sf {C}hemistry with {\sc ADF}}}%
  \bibAnnoteFile{NoStop}{VBB+01}%
\bibitem{C02}%
  \BibitemOpen
  \bibfield{author}{%
  \bibinfo {author} {\bibfnamefont{M.~E.}\ \bibnamefont{Casida}},\ }%
  in\ \emph{\bibinfo {booktitle} {Accurate Description of Low-Lying Molecular
  States and Potential Energy Surfaces}},\ \bibinfo {editor} {edited by\
  \bibinfo {editor} {\bibfnamefont{M.~R.~H.}\ \bibnamefont{Hoffmann}}\ and\
  \bibinfo {editor} {\bibfnamefont{K.~G.}\ \bibnamefont{Dyall}}}\ (\bibinfo
  {publisher} {ACS Press},\ \bibinfo {address} {Washington, D.C.},\ \bibinfo
  {year} {2002})\ p.\ \bibinfo {pages} {199},\ \bibinfo {note} {{\sf {J}acob's
  ladder for time-dependent density-functional theory: {S}ome rungs on the way
  to photochemical heaven}}%
  \bibAnnoteFile{NoStop}{C02}%
\bibitem{D03}%
  \BibitemOpen
  \bibfield{author}{%
  \bibinfo {author} {\bibfnamefont{C.}~\bibnamefont{Daniel}},\ }%
  \bibfield{journal}{%
  \bibinfo {journal} {Coordination Chem. Rev.}\ }%
  \textbf{\bibinfo {volume} {238-239}},\ \bibinfo {pages} {141} (\bibinfo
  {year} {2003}),\ \bibinfo {note} {{\sf {E}lectronic spectroscopy and
  photoreactivity in transition metal complexes}}%
  \bibAnnoteFile{NoStop}{D03}%
\bibitem{MG03}%
  \BibitemOpen
  \bibfield{author}{%
  \bibinfo {author} {\bibfnamefont{M.~A.~L.}\ \bibnamefont{Marques}}\ and\
  \bibinfo {author} {\bibfnamefont{E.~K.~U.}\ \bibnamefont{Gross}},\ }%
  in\ \emph{\bibinfo {booktitle} {A Primer in Density Functional Theory}},\
  \bibinfo {series} {Springer Lecture Notes in Physics}, Vol.\ \bibinfo
  {volume} {620},\ \bibinfo {editor} {edited by\ \bibinfo {editor}
  {\bibfnamefont{C.}~\bibnamefont{Fiolhais}}, \bibinfo {editor}
  {\bibfnamefont{F.}~\bibnamefont{Nogueira}},\ and\ \bibinfo {editor}
  {\bibfnamefont{M.~A.~L.}\ \bibnamefont{Marques}}}\ (\bibinfo {publisher}
  {Springer},\ \bibinfo {year} {2003})\ pp.\ \bibinfo {pages} {144--184},\
  \bibinfo {note} {{\sf Time-dependent density functional theory}}%
  \bibAnnoteFile{NoStop}{MG03}%
\bibitem{MG04}%
  \BibitemOpen
  \bibfield{author}{%
  \bibinfo {author} {\bibfnamefont{M.~A.~L.}\ \bibnamefont{Marques}}\ and\
  \bibinfo {author} {\bibfnamefont{E.~K.~U.}\ \bibnamefont{Gross}},\ }%
  \bibfield{journal}{%
  \bibinfo {journal} {Annu. Rev. Phys. Chem.}\ }%
  \textbf{\bibinfo {volume} {55}},\ \bibinfo {pages} {427} (\bibinfo {year}
  {2004}),\ \bibinfo {note} {{\sf {T}ime-Dependent Density-Functional Theory}}%
  \bibAnnoteFile{NoStop}{MG04}%
\bibitem{BWG05}%
  \BibitemOpen
  \bibfield{author}{%
  \bibinfo {author} {\bibfnamefont{K.}~\bibnamefont{Burke}}, \bibinfo {author}
  {\bibfnamefont{J.}~\bibnamefont{Werschnik}},\ and\ \bibinfo {author}
  {\bibfnamefont{E.~K.~U.}\ \bibnamefont{Gross}},\ }%
  \bibfield{journal}{%
  \bibinfo {journal} {J. Chem. Phys.}\ }%
  \textbf{\bibinfo {volume} {123}},\ \bibinfo {pages} {062206} (\bibinfo {year}
  {2005}),\ \bibinfo {note} {{\sf {T}ime-dependent density functional theory:
  Past, present and future}}%
  \bibAnnoteFile{NoStop}{BWG05}%
\bibitem{DH05}%
  \BibitemOpen
  \bibfield{author}{%
  \bibinfo {author} {\bibfnamefont{A.}~\bibnamefont{Dreuw}}\ and\ \bibinfo
  {author} {\bibfnamefont{M.}~\bibnamefont{Head-Gordon}},\ }%
  \bibfield{journal}{%
  \bibinfo {journal} {Chem. Rev.}\ }%
  \textbf{\bibinfo {volume} {105}},\ \bibinfo {pages} {4009} (\bibinfo {year}
  {2005}),\ \bibinfo {note} {{\sf {S}ingle-reference {\em ab initio} methods
  for the calculation of excited states of large molecules}}%
  \bibAnnoteFile{NoStop}{DH05}%
\bibitem{C08}%
  \BibitemOpen
  \bibfield{author}{%
  \bibinfo {author} {\bibfnamefont{M.~E.}\ \bibnamefont{Casida}},\ }%
  in\ \emph{\bibinfo {booktitle} {Computational Methods in Catalysis and
  Materials Science}},\ \bibinfo {editor} {edited by\ \bibinfo {editor}
  {\bibfnamefont{P.}~\bibnamefont{Sautet}}\ and\ \bibinfo {editor}
  {\bibfnamefont{R.~A.}\ \bibnamefont{van Santen}}}\ (\bibinfo {publisher}
  {Wiley-VCH},\ \bibinfo {address} {Weinheim, Germany},\ \bibinfo {year}
  {2008})\ p.~\bibinfo {pages} {33},\ \bibinfo {note} {{\sf {TDDFT} for Excited
  States}}%
  \bibAnnoteFile{NoStop}{C08}%
\bibitem{C09}%
  \BibitemOpen
  \bibfield{author}{%
  \bibinfo {author} {\bibfnamefont{M.~E.}\ \bibnamefont{Casida}},\ }%
  \bibfield{journal}{%
  \bibinfo {journal} {J. Mol. Struct. (Theochem)}\ }%
  \textbf{\bibinfo {volume} {914}},\ \bibinfo {pages} {3} (\bibinfo {year}
  {2009}),\ \bibinfo {note} {{\sf Review: Time-Dependent Density-Functional
  Theory for Molecules and Molecular Solids}}%
  \bibAnnoteFile{NoStop}{C09}%
\bibitem{CH12}%
  \BibitemOpen
  \bibfield{author}{%
  \bibinfo {author} {\bibfnamefont{M.~E.}\ \bibnamefont{Casida}}\ and\ \bibinfo
  {author} {\bibfnamefont{M.}~\bibnamefont{Huix-Rotllant}},\ }%
  \bibfield{journal}{%
  \bibinfo {journal} {Annu. Rev. Phys. Chem.}\ }%
  \textbf{\bibinfo {volume} {63}},\ \bibinfo {pages} {in press},\ \bibinfo
  {note} {{\sf Progress in Time-Dependent Density-Functional Theory}}%
  \bibAnnoteFile{NoStop}{CH12}%
\bibitem{MS90}%
  \BibitemOpen
  \bibfield{author}{%
  \bibinfo {author} {\bibfnamefont{G.~D.}\ \bibnamefont{Mahan}}\ and\ \bibinfo
  {author} {\bibfnamefont{K.~R.}\ \bibnamefont{Subbaswamy}},\ }%
  \emph{\bibinfo {title} {Local Density Theory of Polarizability}}\ (\bibinfo
  {publisher} {Plenum},\ \bibinfo {address} {New York},\ \bibinfo {year}
  {1990})%
  \bibAnnoteFile{NoStop}{MS90}%
\bibitem{MUN+06}%
  \BibitemOpen
  \emph{\bibinfo {title} {Time-Dependent Density-Functional Theory}},\ edited
  by\ \bibinfo {editor} {\bibfnamefont{M.~A.~L.}\ \bibnamefont{Marques}},
  \bibinfo {editor} {\bibfnamefont{C.}~\bibnamefont{Ullrich}}, \bibinfo
  {editor} {\bibfnamefont{F.}~\bibnamefont{Nogueira}}, \bibinfo {editor}
  {\bibfnamefont{A.}~\bibnamefont{Rubio}},\ and\ \bibinfo {editor}
  {\bibfnamefont{E.~K.~U.}\ \bibnamefont{Gross}},\ \bibinfo {series} {Lecture
  Notes in Physics}, Vol.\ \bibinfo {volume} {706}\ (\bibinfo {publisher}
  {Springer},\ \bibinfo {address} {Berlin},\ \bibinfo {year} {2006})%
  \bibAnnoteFile{NoStop}{MUN+06}%
\bibitem{GMNR11}%
  \BibitemOpen
  \emph{\bibinfo {title} {Fundamentals of Time-Dependent Density-Functional
  Theory}},\ edited by\ \bibinfo {editor}
  {\bibfnamefont{E.}~\bibnamefont{Gross}}, \bibinfo {editor}
  {\bibfnamefont{M.}~\bibnamefont{Marques}}, \bibinfo {editor}
  {\bibfnamefont{F.}~\bibnamefont{Noguiera}},\ and\ \bibinfo {editor}
  {\bibfnamefont{A.}~\bibnamefont{Rubio}},\ Lecture Notes in Physics\ (\bibinfo
  {publisher} {Springer},\ \bibinfo {address} {Berlin},\ \bibinfo {year}
  {2011})\ \bibinfo {note} {in press}%
  \bibAnnoteFile{NoStop}{GMNR11}%
\bibitem{H82}%
  \BibitemOpen
  \bibfield{author}{%
  \bibinfo {author} {\bibfnamefont{R.~C.}\ \bibnamefont{Hilborn}},\ }%
  \bibfield{journal}{%
  \bibinfo {journal} {Am. J. Phys.}\ }%
  \textbf{\bibinfo {volume} {50}},\ \bibinfo {pages} {982} (\bibinfo {year}
  {1982}),\ \bibinfo {note} {{\sf {E}instein coefficients, cross sections, $f$
  values, dipole moments and all that}}%
  \bibAnnoteFile{NoStop}{H82}%
\bibitem{H83}%
  \BibitemOpen
  \bibfield{author}{%
  \bibinfo {author} {\bibfnamefont{R.~C.}\ \bibnamefont{Hilborn}},\ }%
  \bibfield{journal}{%
  \bibinfo {journal} {Am. J. Phys.}\ }%
  \textbf{\bibinfo {volume} {51}},\ \bibinfo {pages} {471} (\bibinfo {year}
  {1983}),\ \bibinfo {note} {{\sf Erratum:``{E}instein coefficients, cross
  sections, $f$ values, dipole moments and all that"[Am. J. Phys. 50, 982
  (1982)]}}%
  \bibAnnoteFile{NoStop}{H83}%
\bibitem{H02}%
  \BibitemOpen
  \bibfield{author}{%
  \bibinfo {author} {\bibfnamefont{R.~C.}\ \bibnamefont{Hilborn}},\ }%
  \enquote{\bibinfo {title} {{\sf Einstein coefficinets, cross sections, $f$
  values, dipole moments, and all that , an updated version of Am. J. Phys. 50,
  982 (1982)}},}\ \bibinfo {howpublished}
  {http://arxiv.org/abs/physics/020202}%
  \bibAnnoteFile{NoStop}{H02}%
\bibitem{BM54a}%
  \BibitemOpen
  \bibfield{author}{%
  \bibinfo {author} {\bibfnamefont{N.~S.}\ \bibnamefont{Bayliss}}\ and\
  \bibinfo {author} {\bibfnamefont{E.~G.}\ \bibnamefont{McRae}},\ }%
  \bibfield{journal}{%
  \bibinfo {journal} {J. Phys. Chem.}\ }%
  \textbf{\bibinfo {volume} {58}},\ \bibinfo {pages} {1002} (\bibinfo {year}
  {1954}),\ \bibinfo {note} {{\sf Solvent Effects in Organic Spectra: Dipole
  Forces and the Franck--Condon Principle}}%
  \bibAnnoteFile{NoStop}{BM54a}%
\bibitem{BM54b}%
  \BibitemOpen
  \bibfield{author}{%
  \bibinfo {author} {\bibfnamefont{N.~S.}\ \bibnamefont{Bayliss}}\ and\
  \bibinfo {author} {\bibfnamefont{E.~G.}\ \bibnamefont{McRae}},\ }%
  \bibfield{journal}{%
  \bibinfo {journal} {J. Phys. Chem.}\ }%
  \textbf{\bibinfo {volume} {58}},\ \bibinfo {pages} {1006} (\bibinfo {year}
  {1954}),\ \bibinfo {note} {{\sf Solvent Effects in the Spectra of Acetone,
  Crotonaldehyde, Nitromethane and Nitrobenzene}}%
  \bibAnnoteFile{NoStop}{BM54b}%
\bibitem{L57}%
  \BibitemOpen
  \bibfield{author}{%
  \bibinfo {author} {\bibfnamefont{E.}~\bibnamefont{Lippert}},\ }%
  \bibfield{journal}{%
  \bibinfo {journal} {Z. Elektrochem.}\ }%
  \textbf{\bibinfo {volume} {61}},\ \bibinfo {pages} {962} (\bibinfo {year}
  {1957}),\ \bibinfo {note} {{\sf Spectroscopic determination of the dipole
  moment of aromatic compounds in the first excited singlet state}}%
  \bibAnnoteFile{NoStop}{L57}%
\bibitem{LP57}%
  \BibitemOpen
  \bibfield{author}{%
  \bibinfo {author} {\bibfnamefont{H.~C.}\ \bibnamefont{Longuet-Higgins}}\ and\
  \bibinfo {author} {\bibfnamefont{J.~A.}\ \bibnamefont{Pople}},\ }%
  \bibfield{journal}{%
  \bibinfo {journal} {J. Chem. Phys.}\ }%
  \textbf{\bibinfo {volume} {27}},\ \bibinfo {pages} {192} (\bibinfo {year}
  {1957}),\ \bibinfo {note} {{\sf Electronic Spectral Shifts of Nonpolar
  Molecules in Nonpolar Solvents}}%
  \bibAnnoteFile{NoStop}{LP57}%
\bibitem{M57}%
  \BibitemOpen
  \bibfield{author}{%
  \bibinfo {author} {\bibfnamefont{E.~G.}\ \bibnamefont{{McRae}}},\ }%
  \bibfield{journal}{%
  \bibinfo {journal} {J. Phys. Chem.}\ }%
  \textbf{\bibinfo {volume} {61}},\ \bibinfo {pages} {562} (\bibinfo {year}
  {1957}),\ \bibinfo {note} {{\sf Theory of Solvent Effects on Molecular
  Electronic Spectra. Frequency Shifts}}%
  \bibAnnoteFile{NoStop}{M57}%
\bibitem{B64}%
  \BibitemOpen
  \bibfield{author}{%
  \bibinfo {author} {\bibfnamefont{S.}~\bibnamefont{Basu}},\ }%
  \bibfield{journal}{%
  \bibinfo {journal} {Adv. Quantum Chem.}\ }%
  \textbf{\bibinfo {volume} {1}},\ \bibinfo {pages} {145} (\bibinfo {year}
  {1964}),\ \bibinfo {note} {{\sf Theory of Solvent Effects on Molecular
  Electronic Spectra}}%
  \bibAnnoteFile{NoStop}{B64}%
\bibitem{BSK76}%
  \BibitemOpen
  \bibfield{author}{%
  \bibinfo {author} {\bibfnamefont{R.~R.}\ \bibnamefont{Birge}}, \bibinfo
  {author} {\bibfnamefont{M.~J.}\ \bibnamefont{Sullivan}},\ and\ \bibinfo
  {author} {\bibfnamefont{B.~E.}\ \bibnamefont{Kohler}},\ }%
  \bibfield{journal}{%
  \bibinfo {journal} {J. Am. Chem. Soc.}\ }%
  \textbf{\bibinfo {volume} {98}},\ \bibinfo {pages} {358} (\bibinfo {year}
  {1976}),\ \bibinfo {note} {{\sf The effect of temperature and solvent
  environment on the conformational stability of 11-cis-retinal}}%
  \bibAnnoteFile{NoStop}{BSK76}%
\bibitem{C34}%
  \BibitemOpen
  \bibfield{author}{%
  \bibinfo {author} {\bibfnamefont{N.~Q.}\ \bibnamefont{Chako}},\ }%
  \bibfield{journal}{%
  \bibinfo {journal} {J. Chem. Phys.}\ }%
  \textbf{\bibinfo {volume} {2}},\ \bibinfo {pages} {644} (\bibinfo {year}
  {1934}),\ \bibinfo {note} {{\sf {A}dsorption of light in organic compounds}}%
  \bibAnnoteFile{NoStop}{C34}%
\bibitem{MR41}%
  \BibitemOpen
  \bibfield{author}{%
  \bibinfo {author} {\bibfnamefont{R.~S.}\ \bibnamefont{Mulliken}}\ and\
  \bibinfo {author} {\bibfnamefont{C.~A.}\ \bibnamefont{Rieke}},\ }%
  \bibfield{journal}{%
  \bibinfo {journal} {Rep. Prog. Phys.}\ }%
  \textbf{\bibinfo {volume} {8}},\ \bibinfo {pages} {231} (\bibinfo {year}
  {1941}),\ \bibinfo {note} {{\sf {M}olecular electronic spectra, dispersion
  and polarization: {T}he theoretical interpretation and computation of
  osciallator strengths and intensities}}%
  \bibAnnoteFile{NoStop}{MR41}%
\bibitem{W64}%
  \BibitemOpen
  \bibfield{author}{%
  \bibinfo {author} {\bibfnamefont{J.}~\bibnamefont{O.~E.~Weigang}},\ }%
  \bibfield{journal}{%
  \bibinfo {journal} {J. Chem. Phys.}\ }%
  \textbf{\bibinfo {volume} {41}},\ \bibinfo {pages} {1435} (\bibinfo {year}
  {1964}),\ \bibinfo {note} {{\sf Solvent Field Corrections for Electric Dipole
  and Rotatory Strengths}}%
  \bibAnnoteFile{NoStop}{W64}%
\bibitem{L66}%
  \BibitemOpen
  \bibfield{author}{%
  \bibinfo {author} {\bibfnamefont{W.}~\bibnamefont{Liptay}},\ }%
  \bibfield{journal}{%
  \bibinfo {journal} {Z. Naturforsch.}\ }%
  \textbf{\bibinfo {volume} {21A}},\ \bibinfo {pages} {1605} (\bibinfo {year}
  {1966}),\ \bibinfo {note} {{\sf Solvent-dependence of the wave number of
  electron bands and physicochemical fundamentals}}%
  \bibAnnoteFile{NoStop}{L66}%
\bibitem{BW68}%
  \BibitemOpen
  \bibfield{author}{%
  \bibinfo {author} {\bibfnamefont{N.~S.}\ \bibnamefont{Bayliss}},\ }%
  \bibfield{journal}{%
  \bibinfo {journal} {Spectrochim. Acta.}\ }%
  \textbf{\bibinfo {volume} {24A}},\ \bibinfo {pages} {551} (\bibinfo {year}
  {1968}),\ \bibinfo {note} {{\sf Solvent effects on the intensities of the
  weak ultraviolet spectra of ketones and nitroparaffins--I}}%
  \bibAnnoteFile{NoStop}{BW68}%
\bibitem{A70}%
  \BibitemOpen
  \bibfield{author}{%
  \bibinfo {author} {\bibfnamefont{T.}~\bibnamefont{Abe}},\ }%
  \bibfield{journal}{%
  \bibinfo {journal} {Bull. Chem. Soc. Jpn.}\ }%
  \textbf{\bibinfo {volume} {43}},\ \bibinfo {pages} {625} (\bibinfo {year}
  {1970}),\ \bibinfo {note} {{\sf Theory of solvent effects on osillator
  strengths for molecular electronic transitions}}%
  \bibAnnoteFile{NoStop}{A70}%
\bibitem{MB80}%
  \BibitemOpen
  \bibfield{author}{%
  \bibinfo {author} {\bibfnamefont{A.~B.}\ \bibnamefont{Meyers}}\ and\ \bibinfo
  {author} {\bibfnamefont{R.~R.}\ \bibnamefont{Birge}},\ }%
  \bibfield{journal}{%
  \bibinfo {journal} {J. Chem. Phys.}\ }%
  \textbf{\bibinfo {volume} {73}},\ \bibinfo {pages} {5314} (\bibinfo {year}
  {1980}),\ \bibinfo {note} {{\sf {T}he effect of solvent environment on
  molecular electronic oscillator strengths}}%
  \bibAnnoteFile{NoStop}{MB80}%
\bibitem{AI86}%
  \BibitemOpen
  \bibfield{author}{%
  \bibinfo {author} {\bibfnamefont{T.}~\bibnamefont{Abe}}\ and\ \bibinfo
  {author} {\bibfnamefont{I.}~\bibnamefont{Iweibo}},\ }%
  \bibfield{journal}{%
  \bibinfo {journal} {Bull. Chem. Soc. Jpn.}\ }%
  \textbf{\bibinfo {volume} {59}},\ \bibinfo {pages} {2381} (\bibinfo {year}
  {1986}),\ \bibinfo {note} {{\sf Solvent effects on the $n-\pi^*$ and
  $\pi-\pi^*$ absorption intensities of some organic molecules}}%
  \bibAnnoteFile{NoStop}{AI86}%
\bibitem{A91}%
  \BibitemOpen
  \bibfield{author}{%
  \bibinfo {author} {\bibfnamefont{J.}~\bibnamefont{Appleqvist}},\ }%
  \bibfield{journal}{%
  \bibinfo {journal} {J. Phys. Chem.}\ }%
  \textbf{\bibinfo {volume} {95}},\ \bibinfo {pages} {3539} (\bibinfo {year}
  {1991}),\ \bibinfo {note} {{\sf Theory of Solvent Effects on the Visible
  Absorption Spectrum of $\beta$-Carotene by a Lattice-Filled Cavity Model}}%
  \bibAnnoteFile{NoStop}{A91}%
\bibitem{HB94}%
  \BibitemOpen
  \bibfield{author}{%
  \bibinfo {author} {\bibfnamefont{T.}~\bibnamefont{Harder}}\ and\ \bibinfo
  {author} {\bibfnamefont{J.}~\bibnamefont{Bendig}},\ }%
  \bibfield{journal}{%
  \bibinfo {journal} {Chem. Phys. Lett.}\ }%
  \textbf{\bibinfo {volume} {228}},\ \bibinfo {pages} {621} (\bibinfo {year}
  {1994}),\ \bibinfo {note} {{\sf Solvent effects on the {UV/VIS} absorption
  spectrum of bis-(dimethylamino)-heptamethinium chloride
  {[(BDH)$^+$Cl$^-$]}}}%
  \bibAnnoteFile{NoStop}{HB94}%
\bibitem{BGK06}%
  \BibitemOpen
  \bibfield{author}{%
  \bibinfo {author} {\bibfnamefont{N.~G.}\ \bibnamefont{Bakhshiev}}, \bibinfo
  {author} {\bibfnamefont{M.~I.}\ \bibnamefont{Gotynyan}},\ and\ \bibinfo
  {author} {\bibfnamefont{A.~V.}\ \bibnamefont{Kirilova}},\ }%
  \bibfield{journal}{%
  \bibinfo {journal} {J. Opt. Technol.}\ }%
  \textbf{\bibinfo {volume} {73}},\ \bibinfo {pages} {666} (\bibinfo {year}
  {2006}),\ \bibinfo {note} {{\sf How the solvent affects the oscillator
  strengths of the intense vibronic absorption bands of polyatomic organic
  molecules}}%
  \bibAnnoteFile{NoStop}{BGK06}%
\bibitem{HH99}%
  \BibitemOpen
  \bibfield{author}{%
  \bibinfo {author} {\bibfnamefont{S.}~\bibnamefont{Hirata}}\ and\ \bibinfo
  {author} {\bibfnamefont{M.}~\bibnamefont{Head-Gordon}},\ }%
  \bibfield{journal}{%
  \bibinfo {journal} {Chem. Phys. Lett.}\ }%
  \textbf{\bibinfo {volume} {314}},\ \bibinfo {pages} {291} (\bibinfo {year}
  {1999}),\ \bibinfo {note} {{\sf {T}ime-dependent density functional theory
  within the {Tamm-Dancoff} approximation}}%
  \bibAnnoteFile{NoStop}{HH99}%
\bibitem{CND11}%
  \BibitemOpen
  \bibfield{author}{%
  \bibinfo {author} {\bibfnamefont{M.~E.}\ \bibnamefont{Casida}}, \bibinfo
  {author} {\bibfnamefont{B.}~\bibnamefont{Natarajan}},\ and\ \bibinfo {author}
  {\bibfnamefont{T.}~\bibnamefont{Deutsch}},\ }%
  \enquote{\bibinfo {title} {Real-time dynamics and conical intersections},}\
  \bibinfo {howpublished} {http://arxiv.org/abs/1102.1849},\ \bibinfo {note}
  {to appear as a chapter in {\em Fundamentals of Time-Dependent
  Density-Functional Theory}, edited by E.K.U. Gross, M. Marques, F. Noguiera,
  and A. Rubio (Springer: in press)}%
  \bibAnnoteFile{NoStop}{CND11}%
\bibitem{GTH96}%
  \BibitemOpen
  \bibfield{author}{%
  \bibinfo {author} {\bibfnamefont{S.}~\bibnamefont{Goedecker}}, \bibinfo
  {author} {\bibfnamefont{M.}~\bibnamefont{Teter}},\ and\ \bibinfo {author}
  {\bibfnamefont{J.}~\bibnamefont{Hutter}},\ }%
  \bibfield{journal}{%
  \bibinfo {journal} {Phys. Rev. B}\ }%
  \textbf{\bibinfo {volume} {54}},\ \bibinfo {pages} {1703} (\bibinfo {year}
  {1996}),\ \bibinfo {note} {{\sf Separable dual-space Gaussian
  pseudopotentials}}%
  \bibAnnoteFile{NoStop}{GTH96}%
\bibitem{CA80}%
  \BibitemOpen
  \bibfield{author}{%
  \bibinfo {author} {\bibfnamefont{D.}~\bibnamefont{Ceperley}}\ and\ \bibinfo
  {author} {\bibfnamefont{B.~J.}\ \bibnamefont{Alder}},\ }%
  \bibfield{journal}{%
  \bibinfo {journal} {Phys. Rev. Lett.}\ }%
  \textbf{\bibinfo {volume} {45}},\ \bibinfo {pages} {566} (\bibinfo {year}
  {1980}),\ \bibinfo {note} {{\sf Ground state of the electron gas by
  stochastic method}}%
  \bibAnnoteFile{NoStop}{CA80}%
\bibitem{DD89}%
  \BibitemOpen
  \bibfield{author}{%
  \bibinfo {author} {\bibfnamefont{G.}~\bibnamefont{Deslauriers}}\ and\
  \bibinfo {author} {\bibfnamefont{S.}~\bibnamefont{Duboc}},\ }%
  \bibfield{journal}{%
  \bibinfo {journal} {Constructive Approx}\ }%
  \textbf{\bibinfo {volume} {5}},\ \bibinfo {pages} {49} (\bibinfo {year}
  {1989}),\ \bibinfo {note} {{\sf {S}ymmetric iterative interpolation
  processes}}%
  \bibAnnoteFile{NoStop}{DD89}%
\bibitem{NG+06}%
  \BibitemOpen
  \bibfield{author}{%
  \bibinfo {author} {\bibfnamefont{A.~I.}\ \bibnamefont{Neelov}}\ and\ \bibinfo
  {author} {\bibfnamefont{S.}~\bibnamefont{Goedecker}},\ }%
  \bibfield{journal}{%
  \bibinfo {journal} {J. Comput. Phys.}\ }%
  \textbf{\bibinfo {volume} {217}},\ \bibinfo {pages} {055501} (\bibinfo {year}
  {2006}),\ \bibinfo {note} {{\sf An efficient numerical quadrature for the
  calculation of the potential energy of wavefunctions expressed in the
  Daubechies wavelet basis}}%
  \bibAnnoteFile{NoStop}{NG+06}%
\bibitem{D75}%
  \BibitemOpen
  \bibfield{author}{%
  \bibinfo {author} {\bibfnamefont{E.~R.}\ \bibnamefont{Davidson}},\ }%
  \bibfield{journal}{%
  \bibinfo {journal} {J. Comput. Phys.}\ }%
  \textbf{\bibinfo {volume} {17}},\ \bibinfo {pages} {87} (\bibinfo {year}
  {1975}),\ \bibinfo {note} {{\sf The iterative calculation of a few of the
  lowest eigenvalues and corresponding eigenvectors of large real-symmetric
  matrices}}%
  \bibAnnoteFile{NoStop}{D75}%
\bibitem{MRD91}%
  \BibitemOpen
  \bibfield{author}{%
  \bibinfo {author} {\bibfnamefont{C.~W.}\ \bibnamefont{Murray}}, \bibinfo
  {author} {\bibfnamefont{S.~C.}\ \bibnamefont{Racine}},\ and\ \bibinfo
  {author} {\bibfnamefont{E.}~\bibnamefont{Davidson}},\ }%
  \bibfield{journal}{%
  \bibinfo {journal} {J. Comput. Phys.}\ }%
  \textbf{\bibinfo {volume} {103}},\ \bibinfo {pages} {382} (\bibinfo {year}
  {1991}),\ \bibinfo {note} {{\sf Improved algorithms for the lowest few
  eigenvalues and eigenvectors of large matrices}}%
  \bibAnnoteFile{NoStop}{MRD91}%
\bibitem{GDN+06}%
  \BibitemOpen
  \bibfield{author}{%
  \bibinfo {author} {\bibfnamefont{L.}~\bibnamefont{Genovese}}, \bibinfo
  {author} {\bibfnamefont{T.}~\bibnamefont{Deutsch}}, \bibinfo {author}
  {\bibfnamefont{A.}~\bibnamefont{Neelov}}, \bibinfo {author}
  {\bibfnamefont{S.}~\bibnamefont{Goedecker}},\ and\ \bibinfo {author}
  {\bibfnamefont{G.}~\bibnamefont{Beylkin}},\ }%
  \bibfield{journal}{%
  \bibinfo {journal} {J. Chem. Phys.}\ }%
  \textbf{\bibinfo {volume} {125}},\ \bibinfo {pages} {074105} (\bibinfo {year}
  {2006}),\ \bibinfo {note} {{\sf Efficient solution of Poisson's equation with
  free boundary conditions}}%
  \bibAnnoteFile{NoStop}{GDN+06}%
\bibitem{GDG07}%
  \BibitemOpen
  \bibfield{author}{%
  \bibinfo {author} {\bibfnamefont{L.}~\bibnamefont{Genovese}}, \bibinfo
  {author} {\bibfnamefont{T.}~\bibnamefont{Deutsch}},\ and\ \bibinfo {author}
  {\bibfnamefont{S.}~\bibnamefont{Goedecker}},\ }%
  \bibfield{journal}{%
  \bibinfo {journal} {J. Chem. Phys.}\ }%
  \textbf{\bibinfo {volume} {127}},\ \bibinfo {pages} {054704} (\bibinfo {year}
  {2007}),\ \bibinfo {note} {{\sf Efficient and accurate three-dimensional
  Poisson solver for surface problems}}%
  \bibAnnoteFile{NoStop}{GDG07}%
\bibitem{GNG+08}%
  \BibitemOpen
  \bibfield{author}{%
  \bibinfo {author} {\bibfnamefont{L.}~\bibnamefont{Genovese}}, \bibinfo
  {author} {\bibfnamefont{A.}~\bibnamefont{Neelov}}, \bibinfo {author}
  {\bibfnamefont{S.}~\bibnamefont{Goedecker}}, \bibinfo {author}
  {\bibfnamefont{T.}~\bibnamefont{Deutsch}}, \bibinfo {author}
  {\bibfnamefont{S.~A.}\ \bibnamefont{Ghasemi}}, \bibinfo {author}
  {\bibfnamefont{A.}~\bibnamefont{Willand}}, \bibinfo {author}
  {\bibfnamefont{D.}~\bibnamefont{Caliste}}, \bibinfo {author}
  {\bibfnamefont{O.}~\bibnamefont{Zilberberg}}, \bibinfo {author}
  {\bibfnamefont{M.}~\bibnamefont{Rayson}}, \bibinfo {author}
  {\bibfnamefont{A.}~\bibnamefont{Bergman}},\ and\ \bibinfo {author}
  {\bibfnamefont{R.}~\bibnamefont{Schneider}},\ }%
  \bibfield{journal}{%
  \bibinfo {journal} {J. Chem. Phys.}\ }%
  \textbf{\bibinfo {volume} {129}},\ \bibinfo {pages} {014109} (\bibinfo {year}
  {2008}),\ \bibinfo {note} {{\sf {D}ebauchies wavelets as a basis set for
  density functional pseudopotential calculations}}%
  \bibAnnoteFile{NoStop}{GNG+08}%
\bibitem{AGGNG09}%
  \BibitemOpen
  \bibfield{author}{%
  \bibinfo {author} {\bibfnamefont{M.}~\bibnamefont{Amsler}}, \bibinfo {author}
  {\bibfnamefont{S.~A.}\ \bibnamefont{Ghasemi}}, \bibinfo {author}
  {\bibfnamefont{S.}~\bibnamefont{Goedecker}}, \bibinfo {author}
  {\bibfnamefont{A.}~\bibnamefont{Neelov}},\ and\ \bibinfo {author}
  {\bibfnamefont{L.}~\bibnamefont{Genovese}},\ }%
  \bibfield{journal}{%
  \bibinfo {journal} {Nanotechnology}\ }%
  \textbf{\bibinfo {volume} {20}},\ \bibinfo {pages} {445301} (\bibinfo {year}
  {2009}),\ \bibinfo {note} {{\sf {A}dsorption of small {NaCl} clusters on
  surfaces of silicon nanostructures}}%
  \bibAnnoteFile{NoStop}{AGGNG09}%
\bibitem{GODMNG09}%
  \BibitemOpen
  \bibfield{author}{%
  \bibinfo {author} {\bibfnamefont{L.}~\bibnamefont{Genovese}}, \bibinfo
  {author} {\bibfnamefont{M.}~\bibnamefont{Ospici}}, \bibinfo {author}
  {\bibfnamefont{T.}~\bibnamefont{Deutsch}}, \bibinfo {author}
  {\bibfnamefont{J.-F.}\ \bibnamefont{Méhaut}}, \bibinfo {author}
  {\bibfnamefont{A.}~\bibnamefont{Neelov}},\ and\ \bibinfo {author}
  {\bibfnamefont{S.}~\bibnamefont{Goedecker}},\ }%
  \bibfield{journal}{%
  \bibinfo {journal} {J. Chem. Phys.}\ }%
  \textbf{\bibinfo {volume} {131}},\ \bibinfo {pages} {034103} (\bibinfo {year}
  {2009}),\ \bibinfo {note} {{\sf Density Functional Theory calculation on
  many-cores hybrid CPU-GPU architectures}}%
  \bibAnnoteFile{NoStop}{GODMNG09}%
\bibitem{GVO+11}%
  \BibitemOpen
  \bibfield{author}{%
  \bibinfo {author} {\bibfnamefont{L.}~\bibnamefont{Genovese}}, \bibinfo
  {author} {\bibfnamefont{B.}~\bibnamefont{Videau}}, \bibinfo {author}
  {\bibfnamefont{M.}~\bibnamefont{Ospici}}, \bibinfo {author}
  {\bibfnamefont{T.}~\bibnamefont{Deutsch}}, \bibinfo {author}
  {\bibfnamefont{S.}~\bibnamefont{Goedecker}},\ and\ \bibinfo {author}
  {\bibfnamefont{J.-F.}\ \bibnamefont{Méhaut}},\ }%
  \bibfield{journal}{%
  \bibinfo {journal} {Comptes Rendus M\'{e}canique}\ }%
  \textbf{\bibinfo {volume} {339}},\ \bibinfo {pages} {149} (\bibinfo {year}
  {2011}),\ \bibinfo {note} {{\sf Daubechies wavelets for high performance
  electronic structure calculations}}%
  \bibAnnoteFile{NoStop}{GVO+11}%
\bibitem{F96}%
  \BibitemOpen
  \bibfield{author}{%
  \bibinfo {author} {\bibfnamefont{D.}~\bibnamefont{Feller}},\ }%
  \bibfield{journal}{%
  \bibinfo {journal} {J. Comp. Chem.}\ }%
  \textbf{\bibinfo {volume} {17}},\ \bibinfo {pages} {1571} (\bibinfo {year}
  {1996}),\ \bibinfo {note} {{\sf The role of databases in support of
  computational chemistry calculations}}%
  \bibAnnoteFile{NoStop}{F96}%
\bibitem{SDESGCW07}%
  \BibitemOpen
  \bibfield{author}{%
  \bibinfo {author} {\bibfnamefont{K.~L.}\ \bibnamefont{Schuchardt}}, \bibinfo
  {author} {\bibfnamefont{B.~T.}\ \bibnamefont{Didier}}, \bibinfo {author}
  {\bibfnamefont{T.}~\bibnamefont{Elsethagen}}, \bibinfo {author}
  {\bibfnamefont{L.}~\bibnamefont{Sun}}, \bibinfo {author}
  {\bibfnamefont{V.}~\bibnamefont{Gurumoorthi}}, \bibinfo {author}
  {\bibfnamefont{J.}~\bibnamefont{Chase}}, \bibinfo {author}
  {\bibfnamefont{J.}~\bibnamefont{Li}},\ and\ \bibinfo {author}
  {\bibfnamefont{T.~L.}\ \bibnamefont{Windus}},\ }%
  \bibfield{journal}{%
  \bibinfo {journal} {J. Chem. Inf. Model.}\ }%
  \textbf{\bibinfo {volume} {47}},\ \bibinfo {pages} {1045} (\bibinfo {year}
  {2007}),\ \bibinfo {note} {{\sf Basis set exchange: A community database for
  computational sciences}}%
  \bibAnnoteFile{NoStop}{SDESGCW07}%
\bibitem{CGG+00}%
  \BibitemOpen
  \bibfield{author}{%
  \bibinfo {author} {\bibfnamefont{M.~E.}\ \bibnamefont{Casida}}, \bibinfo
  {author} {\bibfnamefont{F.}~\bibnamefont{Gutierrez}}, \bibinfo {author}
  {\bibfnamefont{J.}~\bibnamefont{Guan}}, \bibinfo {author}
  {\bibfnamefont{F.}~\bibnamefont{Gadea}}, \bibinfo {author}
  {\bibfnamefont{D.~R.}\ \bibnamefont{Salahub}},\ and\ \bibinfo {author}
  {\bibfnamefont{J.}~\bibnamefont{Daudey}},\ }%
  \bibfield{journal}{%
  \bibinfo {journal} {J. Chem. Phys.}\ }%
  \textbf{\bibinfo {volume} {113}},\ \bibinfo {pages} {7062} (\bibinfo {year}
  {2000}),\ \bibinfo {note} {{\sf Charge-transfer correction for improved
  time-dependent local density approximation excited-state potential energy
  curves: {A}nalysis within the two-level model with illustration for {H$_2$}
  and {LiH}}}%
  \bibAnnoteFile{NoStop}{CGG+00}%
\bibitem{abinit-tddft}%
  \BibitemOpen
  \bibinfo {howpublished}
  {\url{http://www.abinit.org/documentation/helpfiles/for-v5.8/tutorial/lesson%
_tddft.html}}%
  \bibAnnoteFile{NoStop}{abinit-tddft}%
\bibitem{BK90}%
  \BibitemOpen
  \bibfield{author}{%
  \bibinfo {author} {\bibfnamefont{S.~B.}\ \bibnamefont{Ben-Shlomo}}\ and\
  \bibinfo {author} {\bibfnamefont{U.}~\bibnamefont{Kaldor}},\ }%
  \bibfield{journal}{%
  \bibinfo {journal} {J. Chem. Phys.}\ }%
  \textbf{\bibinfo {volume} {92}},\ \bibinfo {pages} {3680} (\bibinfo {year}
  {1990}),\ \bibinfo {note} {{\sf N$_2$ excitations below 15 eV by the
  multireference coupled -cluster method}}%
  \bibAnnoteFile{NoStop}{BK90}%
\bibitem{abinit}%
  \BibitemOpen
  \bibfield{author}{%
  \bibinfo {author} {\bibfnamefont{X.}~\bibnamefont{Gonze}}, \bibinfo {author}
  {\bibfnamefont{J.}~\bibnamefont{Beuken}}, \bibinfo {author}
  {\bibfnamefont{R.}~\bibnamefont{Caracas}}, \bibinfo {author}
  {\bibfnamefont{F.}~\bibnamefont{Detraux}}, \bibinfo {author}
  {\bibfnamefont{M.}~\bibnamefont{Fuchs}}, \bibinfo {author}
  {\bibfnamefont{G.}~\bibnamefont{Rignanese}}, \bibinfo {author}
  {\bibfnamefont{L.}~\bibnamefont{Sindic}}, \bibinfo {author}
  {\bibfnamefont{M.}~\bibnamefont{Verstraete}}, \bibinfo {author}
  {\bibfnamefont{G.}~\bibnamefont{Zerah}},\ and\ \bibinfo {author}
  {\bibfnamefont{F.}~\bibnamefont{Jollet}},\ }%
  \bibfield{journal}{%
  \bibinfo {journal} {Comput. Mat. Sci.}\ }%
  \textbf{\bibinfo {volume} {25}},\ \bibinfo {pages} {478} (\bibinfo {year}
  {2002}),\ \bibinfo {note} {{\sf First-principles computation of material
  properties: the ABINIT software project}}%
  \bibAnnoteFile{NoStop}{abinit}%
\bibitem{PW92}%
  \BibitemOpen
  \bibfield{author}{%
  \bibinfo {author} {\bibfnamefont{J.~P.}\ \bibnamefont{Perdew}}\ and\ \bibinfo
  {author} {\bibfnamefont{Y.}~\bibnamefont{Wang}},\ }%
  \bibfield{journal}{%
  \bibinfo {journal} {Phys. Rev. B}\ }%
  \textbf{\bibinfo {volume} {45}},\ \bibinfo {pages} {13244} (\bibinfo {year}
  {1992}),\ \bibinfo {note} {{\sf Accurate and simple analytic representation
  of the electron-gas correlation energy}}%
  \bibAnnoteFile{NoStop}{PW92}%
\bibitem{GR03}%
  \BibitemOpen
  \bibfield{author}{%
  \bibinfo {author} {\bibfnamefont{Z.~R.}\ \bibnamefont{Grabowski}}, \bibinfo
  {author} {\bibfnamefont{K.}~\bibnamefont{Rotkiewicz}},\ and\ \bibinfo
  {author} {\bibfnamefont{W.}~\bibnamefont{Rettig}},\ }%
  \bibfield{journal}{%
  \bibinfo {journal} {Chem. Rev.}\ }%
  \textbf{\bibinfo {volume} {103}},\ \bibinfo {pages} {3899} (\bibinfo {year}
  {2003}),\ \bibinfo {note} {{\sf Structural Changes Accompanying
  Intramolecular Electron Transfer: Focus on Twisted Intramolecular
  Charge-Transfer States and Structures}}%
  \bibAnnoteFile{NoStop}{GR03}%
\bibitem{u62}%
  \BibitemOpen
  \bibfield{author}{%
  \bibinfo {author} {\bibfnamefont{I.}~\bibnamefont{Ugi}},\ }%
  \bibfield{journal}{%
  \bibinfo {journal} {Ang. Chemie Int.}\ }%
  \textbf{\bibinfo {volume} {1}},\ \bibinfo {pages} {8} (\bibinfo {year}
  {1962}),\ \bibinfo {note} {{\sf The $\alpha$-Addition of Immonium Ions and
  Anions to Isonitriles Accompanied by Secondary Reactions}}%
  \bibAnnoteFile{NoStop}{u62}%
\bibitem{ACGG93}%
  \BibitemOpen
  \bibfield{author}{%
  \bibinfo {author} {\bibfnamefont{A.}~\bibnamefont{Altomare}}, \bibinfo
  {author} {\bibfnamefont{G.}~\bibnamefont{Cascarano}}, \bibinfo {author}
  {\bibfnamefont{C.}~\bibnamefont{Giacovazzo}},\ and\ \bibinfo {author}
  {\bibfnamefont{A.}~\bibnamefont{Guagliardi}},\ }%
  \bibfield{journal}{%
  \bibinfo {journal} {J. Appl. Cryst.}\ }%
  \textbf{\bibinfo {volume} {26}},\ \bibinfo {pages} {343} (\bibinfo {year}
  {1993}),\ \bibinfo {note} {{\sf {C}ompletion and refinement of crystal
  structures with {\em SIR92}}}%
  \bibAnnoteFile{NoStop}{ACGG93}%
\bibitem{TeXsan}%
  \BibitemOpen
  \bibinfo {note} {{\sf Molecular Structure Corporation TeXsan. Single Crystal
  Structure Analysis Software. Version 1.7. MSC, 3200 Research Forest Drive,
  The Woodlands, TX 77381, USA. 1992-1997.}}%
  \bibAnnoteFile{Stop}{TeXsan}%
\bibitem{pablo}%
  \BibitemOpen
  \bibinfo {howpublished}
  {\url{http://sites.google.com/site/markcasida/tddft}}%
  \bibAnnoteFile{NoStop}{pablo}%
\bibitem{ORR02}%
  \BibitemOpen
  \bibfield{author}{%
  \bibinfo {author} {\bibfnamefont{G.}~\bibnamefont{Onida}}, \bibinfo {author}
  {\bibfnamefont{L.}~\bibnamefont{Reining}},\ and\ \bibinfo {author}
  {\bibfnamefont{A.}~\bibnamefont{Rubio}},\ }%
  \bibfield{journal}{%
  \bibinfo {journal} {Rev. Mod. Phys.}\ }%
  \textbf{\bibinfo {volume} {74}},\ \bibinfo {pages} {601} (\bibinfo {year}
  {2002}),\ \bibinfo {note} {{\sf {E}lectronic excitations:
  {Density-functional} versus many-body {Green's}-function approaches}}%
  \bibAnnoteFile{NoStop}{ORR02}%
\end{thebibliography}%
\end{document}